\documentclass[aps,prd,twocolumn,superscriptaddress,showpacs,nofootinbib,preprintnumbers,10pt]{revtex4-2}

\usepackage{style}

\begin{document}

\preprint{FERMILAB-PUB-24-0727-ETD-PPD-T}

\title{Listening For New Physics With Quantum Acoustics}
\date{\today}

\author{Ryan Linehan\,\orcidlink{0000-0003-2785-018X}}
\thanks{Corresponding author: \href{linehan3@fnal.gov}{linehan3@fnal.gov}}
\affiliation{Fermi National Accelerator Laboratory, Batavia, IL 60510, USA}

\author{Tanner Trickle\,\orcidlink{0000-0003-1371-4988}}
\affiliation{Theoretical Physics Division, Fermi National Accelerator Laboratory, Batavia, IL 60510, USA}

\author{Christopher R. Conner\,\orcidlink{0000-0002-6875-2298}}
\affiliation{Pritzker School of Molecular Engineering, University of Chicago, Chicago, IL 60637, USA}

\author{Sohitri~Ghosh\,\orcidlink{0000-0003-0082-7772}}
\affiliation{Theoretical Physics Division, Fermi National Accelerator Laboratory, Batavia, IL 60510, USA}

\author{Tongyan Lin\,\orcidlink{0000-0003-4969-3285}}
\affiliation{Department of Physics, University of California, San Diego, CA 92093, USA }

\author{Mukul Sholapurkar\,\orcidlink{0000-0002-2417-0616}}
\affiliation{Department of Physics, University of California, San Diego, CA 92093, USA }
\affiliation{Institute of High Energy Physics, Austrian Academy of Sciences, Dominikanerbastei 16, 1010 Vienna, Austria}

\author{Andrew N. Cleland\,\orcidlink{0000-0003-4981-4294}}
\affiliation{Pritzker School of Molecular Engineering, University of Chicago, Chicago, IL 60637, USA}
\affiliation{Center for Molecular Engineering and Material Science Division, Argonne National Laboratory, Lemont, Illinois 60439, USA}

\begin{abstract}
We present a novel application of a qubit-coupled phonon detector to search for new physics, e.g., ultralight dark matter (DM) and high-frequency gravitational waves. The detector, motivated by recent advances in quantum acoustics, is composed of superconducting transmon qubits coupled to high-overtone bulk acoustic resonators ($h$BARs) and operates in the GHz - 10 GHz frequency range. New physics can excite $\mathcal{O}(10 \, \mu \text{eV})$ phonons within the $h$BAR, which are then converted to qubit excitations via a transducer. We detail the design, operation, backgrounds, and expected sensitivity of a prototype detector, as well as a next-generation detector optimized for new physics signals. We find that a future detector can complement current haloscope experiments in the search for both dark photon DM and high-frequency gravitational waves. Lastly we comment on such a detector's ability to operate as a $10 \, \mu\text{eV}$ threshold athermal phonon sensor for sub-GeV DM detection.
\end{abstract}
\maketitle

\section{Introduction}
\label{sec:Introduction}

Phonons, the quanta of vibrational modes, have proven to be an instrumental sensing channel in detectors employed for a variety of scientific applications, including dark matter (DM) and gravitational wave (GW) detection~\cite{Weber:1960zz,Forward:1971mel,Lobo:1999ns,Hamilton:1989nb,Johnson:1993cr,Aguiar:2010kn,DaSilvaCosta:2014trv,Harry:1996gh,Gottardi:2007zn,Goryachev:2014yra,Goryachev:2021zzn,Tobar:2023ksi,Campbell_2023,SuperCDMS:2024yiv}. In recent years, there has been recognition that new phonon sensing capabilities may further extend sensitivity to these phenomena, enabling probes of a much wider range of DM theories~\cite{Schutz:2016tid,Knapen:2017ekk,Carney:2020xol} as well as novel sources of GWs~\cite{Goryachev:2014yra,Kahn:2023mrj}. This recognition has come hand in hand with a dramatic expansion in theories regarding the fundamental constituents of DM and their production in the early universe~\cite{Knapen:2017xzo,Lin:2019uvt,Zurek:2024qfm}. In many of these theories, the DM candidate is lighter than the prototypical weakly-interacting massive particle, and has vastly different phenomenology. Searching for these DM candidates leads to new signatures and requires new techniques for their detection. 

One example comes from the use of resonant mass detectors~\cite{Weber:1960zz,Forward:1971mel,Hamilton:1989nb,Johnson:1993cr,Harry:1996gh,Lobo:1999ns,Gottardi:2007zn,Aguiar:2010kn,DaSilvaCosta:2014trv} and bulk acoustic wave resonators~\cite{Goryachev:2014yra,Goryachev:2021zzn,Tobar:2023ksi,Campbell_2023} for GWs in the kHz - GHz frequency range. These same devices can probe ultralight DM, which behaves as a classical field oscillating at a frequency equal to its mass, $\omega \approx m_\text{DM}$~\cite{Antypas:2022asj}. Such DM could have the form of a light modulus, leading to an oscillating size for a solid and thus exciting phonon modes. It was shown that for moduli DM, resonant mass and bulk acoustic wave GW detectors can be a sensitive probe in the mass range $10^{-12} \, \text{eV} \lesssim m_\text{DM} \lesssim 10^{-6} \, \text{eV}$~\cite{Arvanitaki:2015iga}.

Phonon sensing in the THz - 100 THz frequency regime can also be used to extend direct detection of DM to lower masses than those accessible by traditional direct detection experiments~\cite{CDMS-II:2006zpy,XENON:2018voc,LZ:2022lsv}.
While phonons are often produced as secondaries in detectors aiming to sense DM with mass greater than a GeV, sub-GeV DM candidates can directly produce phonon excitations through scattering and absorption processes in crystal~\cite{Knapen:2017ekk,Griffin:2018bjn,Trickle:2019nya,Campbell-Deem:2019hdx,Cox:2019cod,Kurinsky:2019pgb,Griffin:2019mvc,Trickle:2020oki,Griffin:2020lgd,Mitridate:2020kly,Coskuner:2021qxo,Knapen:2021bwg,Campbell-Deem:2022fqm,Taufertshofer:2023rgq,Mitridate:2023izi,Lin:2023slv} and superfluid~\cite{Schutz:2016tid,Knapen:2016cue,Acanfora:2019con,Caputo:2019cyg,Caputo:2019xum,Caputo:2020sys,Baym:2020uos,Matchev:2021fuw,You:2022pyn} targets. These phonons could be detected in planned experiments based on single-phonon sensing (e.g., TESSERACT~\cite{Chang2020}) with energy thresholds of $\mathcal{O}(\text{meV})$ ( corresponding to $\mathcal{O}(\text{THz})$ frequency thresholds). Achieving an $\mathcal{O}(\text{meV})$ energy threshold would allow sensitivity down to $m_\text{DM} \gtrsim \text{keV}$ via a scattering event (an important milestone since $m_\text{DM} \sim \text{keV}$ is the lightest fermionic DM is allowed to be~\cite{Tremaine:1979we}), and $m_\text{DM} \gtrsim \text{meV}$ via an absorption event. Many of the detector technologies aiming to reach such thresholds rely on superconducting ``pairbreaking'' sensors such as transition edge sensors, kinetic inductance detectors, superconducting qubits, and quantum capacitance detectors~\cite{McEwen,Harrington:2024iqm,Temples:2024ntv,Fink_2020,Watkins:2022wlz,RenQET,CPD:2020xvi,Linehan:2024niv,ramanathan2024quantumparitydetectorsqubit,Fink:2023tvb}. These derive their competitive sensitivity from the small amount of energy needed to break a single Cooper pair and produce Bogoliubov quasiparticles that can be sensed. However this also fundamentally limits their energy threshold to the $\mathcal{O}(\text{meV})$ energy scale.

Detectors capable of sensing THz phonons are also sensitive to high-frequency GWs~\cite{Kahn:2023mrj}. Such high-frequency GWs are a unique probe of new physics, since there are no known conventional astrophysical sources of GWs with frequencies above $\sim \text{MHz}$~\cite{Aggarwal:2020olq}. Potential sources of GWs above $\sim \text{MHz}$ include a stochastic gravitational wave background from the early universe, and coherent sources from nearby exotic objects, e.g., superradiance of a light field around a black hole~\cite{Arvanitaki:2010sy,Arvanitaki:2014wva,Brito:2015oca,Baryakhtar:2020gao}, or the inspiral of light compact objects~\cite{Maggiore:2007ulw}. There has been significant recent interest in understanding how existing detectors for ultralight DM, including those based on electromagnetic signals, can be used to probe these high-frequency GWs~\cite{Aggarwal:2020olq,Domcke:2023qle}. 

New technologies for phonon detection have the potential to explore new regimes in both DM and GW physics. The field of quantum acoustics, which demonstrates control of individual GHz-frequency phonons, offers one promising direction~\cite{satzingersaws, bienfaitsaws,dumursaws,qiaosaws}. An architecture that enables this is the qubit-coupled high-overtone bulk acoustic resonator ($h$BAR)~\cite{hBAR_masters,Chu_2017,GokhaleHBARs,KervinenPlanar,von2022parity}, 
where the $h$BAR is a crystal substrate designed to host GHz-scale phonon modes. The $h$BAR is coupled to a superconducting transmon qubit via a piezoelectric transducer, with manipulation and sensing of phonon modes performed via qubit operations. 

In this work we propose using a qubit-coupled $h$BAR (qc-$h$BAR) device as a single phonon detector in a new frequency range: $1 \, \text{GHz} \lesssim \omega / 2 \pi \lesssim 10 \, \text{GHz}$ (or energy range $4 \, \mu\text{eV} \lesssim \omega \lesssim 40 \, \mu\text{eV}$). This device architecture is sensitive to any new physics signal which can generate GHz - 10 GHz frequency phonons inside the $h$BAR, where the frequency boundaries are set by the operating frequencies of typical superconducting qubits, and by thermal excitations when operating at $\mathcal{O}(10 \, \text{mK})$ temperatures. This design is complementary to other qubit-based detectors: rather than coupling a qubit with an electromagnetic cavity~\cite{Dixit:2020ymh} or directly using the qubit as the target~\cite{Chen:2022quj}, it couples a readout qubit to a solid-state target, enabling sensitivity to new physics signals which couple directly to matter (as opposed to those which only couple electromagnetically). Moreover this design also complements pairbreaking sensor architectures since it does not rely on Bogoliubov quasiparticles for its phonon sensing, and therefore does not face the same fundamental sensor thresholds.

This paper is organized as follows. In Sec.~\ref{sec:DeviceDesignAndOperation} we present the detector design, the phonon modes it hosts, and its operation. In Sec.~\ref{sec:DevicePerformance} we detail the expected performance of such a detector, first deriving the $h$BAR phonon-qubit swap efficiency and then discussing background estimates. Sec.~\ref{sec:DarkMatterSearchModes} explores the sensitivity of the qc-$h$BAR to signals from dark photon DM and high-frequency GWs. Lastly, in Sec.~\ref{sec:Conclusions} we conclude and discuss potential sensitivity to other new physics signals, e.g., light, sub-GeV DM scattering. Throughout we work in natural units with $c = \hbar = k_\text{B} = 1$.

\begin{figure*}[htbp]
\centering
\includegraphics[width=\linewidth]{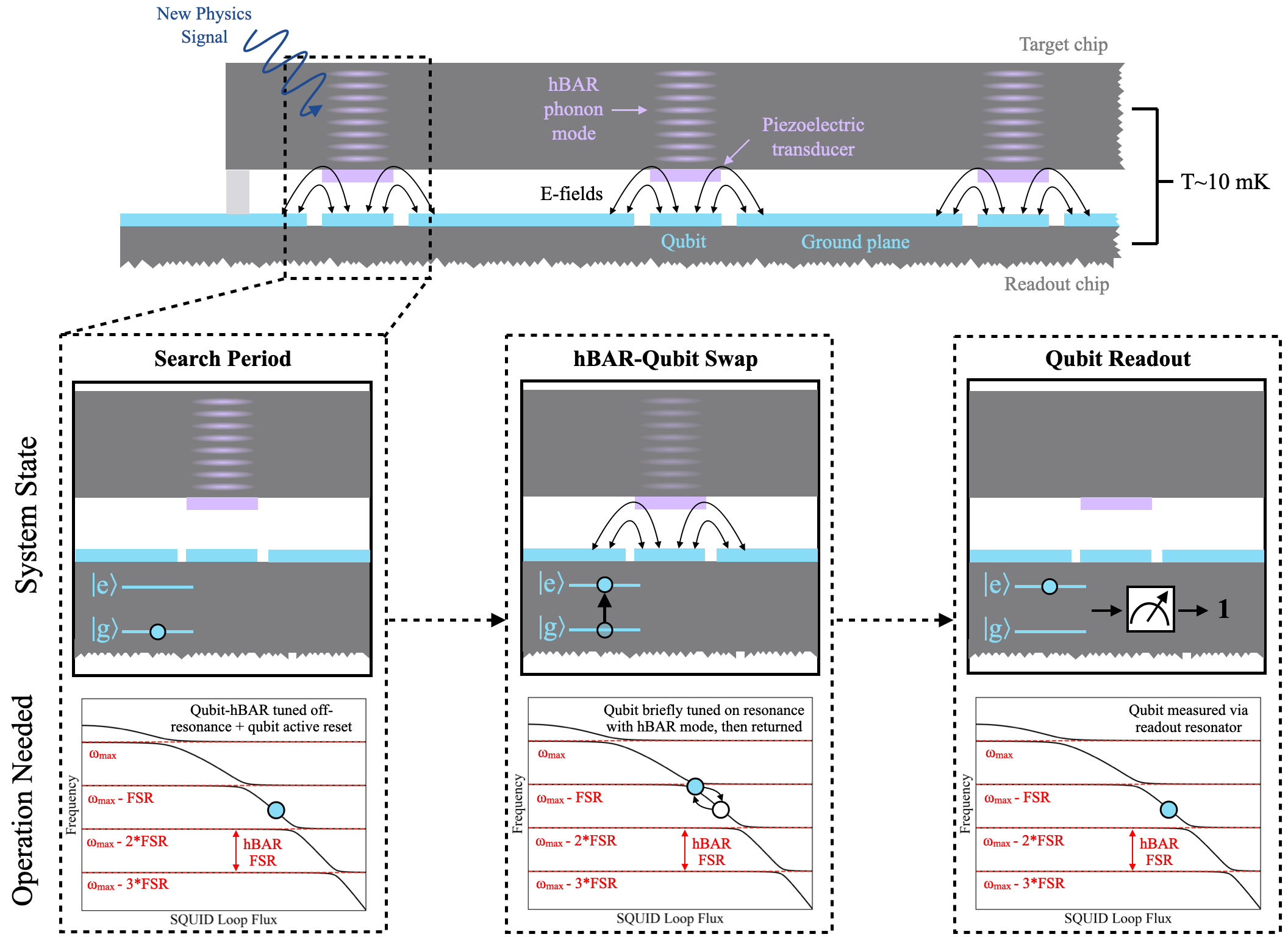}
\caption{\textbf{Top:} Diagram of a generic flip-chip architecture with several qubit-coupled \hbarName{} (qc-$h$BAR) devices. Here we loosely portray a 2D transmon architecture, in which the qubit and readout resonator are both etched into the ground plane on the readout chip. \textbf{Bottom}: For a single qc-$h$BAR, the three outlined panels present a simplified illustration of the detection scheme in which new physics creates a single low-energy phonon (left, labeled ``Search Period"), that phonon is swapped into the qubit (middle, labeled ``$h$BAR-Qubit Swap"), and that qubit is read out (right, labeled ``Qubit Readout"). Each panel details the system state, and operations needed to achieve that state, for each stage of the detection scheme. The bottom figure of each panel shows the frequency (energy) eigenstates of an example \hbarName{} coupled to a flux-tunable superconducting qubit (black), as a function of the flux through the SQUID loop. Horizontal red dashed lines indicate the (uncoupled) \hbarName{} mode frequencies. Avoided level crossings from hybridization of the \hbarName{} and qubit systems occur when the qubit is tuned onto resonance with an \hbarName{} phonon mode frequency. The blue dot represents the frequency to which the qubit must be tuned to accomplish each step.}
\label{fig:DeviceDesignDiagram}
\end{figure*} 

\section{Device Design and Operation}
\label{sec:DeviceDesignAndOperation}

The qc-$h$BAR, shown in Fig.~\ref{fig:DeviceDesignDiagram}, is composed of two components: the $h$BAR in an upper ``target'' chip and the superconducting qubit with an extended qubit island on a bottom ``readout'' chip. These two are assembled in a flip-chip geometry and mounted to the \(10\)~mK stage of a dilution refrigerator. The qubit and $h$BAR are coupled by a piezoelectric transducer (PT) on the underside of the target chip which converts phonons, or strain fluctuations, in the $h$BAR into fluctuations in the electric field of the qubit. The state of the qubit can then be read out dispersively through a separate superconducting microwave resonator~\cite{BlaisCQED,SchusterThesis} (not shown in Fig.~\ref{fig:DeviceDesignDiagram}). In this section, we present a simplified view of the system architecture (Sec.~\ref{subsec:SystemArchitecture}), discuss the phonon modes in the $h$BAR in detail (Sec.~\ref{subsec:PhononModes}), and lastly discuss a nominal readout scheme (Sec.~\ref{subsec:PhononSensingReadoutScheme}) that will facilitate searches for new physics creating the roughly GHz frequency phonons in the $h$BAR.

\subsection{System Architecture}
\label{subsec:SystemArchitecture}

The qc-$h$BAR transfers phonons generated by some new physics signal in the $h$BAR into qubit excitations via the PT. Considering only an individual phonon mode, a simple model describing single-quanta dynamics of the coupled phonon-qubit system is,
\begin{equation}                 
    \label{eq:FullHamiltonian}
    H = \omega_\text{q} \, a^{\dag}a + \omega_{n} \, b^{\dag}b + g \, (a^{\dag}b + ab^{\dag}) \, ,
\end{equation}
where $\omega_\text{q}$ is the electromagnetic energy of the qubit mode, $\omega_n$ is the mechanical energy of the $h$BAR phonon mode, and $g$ is a coupling coefficient determined by the physics of the PT-qubit coupling (derived in detail in Sec.~\ref{subsec:phonon_qubit_swap_efficiency}). $a^\dagger, a$, and $b^\dagger, b$ are the raising and lowering operators of the qubit and $h$BAR phonon modes, respectively. 

The first term in Eq.~\eqref{eq:FullHamiltonian} arises due to the electromagnetic energy stored by a superconducting qubit. Physically, such qubits are electrical circuit elements patterned into a thin film of superconductor that lies on a larger substrate. The circuit design chosen here is a capacitor in parallel with a Josephson junction~\cite{XmonBarends}, which can generate oscillating electromagnetic fields between a central qubit ``island'' and a ground plane, as shown in Fig.~\ref{fig:DeviceDesignDiagram}.\footnote{This design is known as a ``2D" qubit. In a ``3D" qubit, the oscillating electromagnetic field is generated between a pair of capacitor pads in a 3D microwave cavity. Both 2D~\cite{CrumpHBARs,KervinenPlanar} and 3D qubits~\cite{Chu_2017,von2022parity} have been demonstrated in integration with $h$BAR devices.} Functionally, such a circuit forms an anharmonic oscillator, in which the energy spacing \(\omega_{e}-\omega_{g}\) between the ground state \(\ket{g}\) and first excited state \(\ket{e}=a^\dagger \ket{g}\) differs from the spacing \(\omega_{f}-\omega_{e}\) between \(\ket{e}\) and the second excited state $\ket{f}$. This anharmonicity, defined as $\eta \equiv ((\omega_f - \omega_e) - ( \omega_e - \omega_g ))/2\pi$, is large enough to enable controlled manipulations of the system within the subspace spanned by $\ket{g}$ and $\ket{e}$ and allows the qubit to be approximated as a two level system with a transition frequency \(\omega_\text{q} \equiv \omega_{e}-\omega_{g}\). Typically, these qubits are made with transition frequencies in the $4 \, \text{GHz} \lesssim \omega_\text{q}/2\pi \lesssim 8 \, \text{GHz}$ range and $\eta\approx-200$~MHz. In this work, we consider transmon qubits with $1\,\text{GHz} \lesssim \omega_\text{q}/2\pi \lesssim 10 \, \text{GHz}$. The upper end of this range is marginally limited by conventional readout electronics, and transmons have been operated at frequencies as high as 20~GHz~\cite{AnferovQubitsAbove20GHz}. The low end of this range is limited by thermal noise, but \(\simeq1\)~GHz transmon frequencies may still be achieved by lowering the Josephson energy and anharmonicity, at the cost of needing slower qubit operations to keep the qubit in the subspace spanned by $\ket{e}$ and $\ket{g}$. For the specific setup described here, we consider a transmon qubit with a SQUID loop forming the Josephson junction element, so that its frequency \(\omega_\text{q}\) can be tuned in situ with an applied magnetic flux bias.

The second term in Eq.~\eqref{eq:FullHamiltonian} describes a phonon in an $h$BAR mode. This is the fundamental sensing element of the system and acts as a bulk acoustic resonator. Boundary conditions and surface features are important when engineering the phonon modes within a bulk acoustic resonator~\cite{10.1121/1.1912835,Goryachev:2014yra,hBAR_masters} since they can lead to spatially confined phonon modes. As shown in Fig.~\ref{fig:DeviceDesignDiagram}, there are localized phonon modes underneath the PT ``bump''. It is these phonons modes that we focus on as they give rise to the largest strains within the PT and thus the greatest chance to be converted to a qubit excitation. Underneath the PT the phonon mode function has the approximate form, 
\begin{align}
\label{eq:simple_mode_function}
    \vec{u} \propto  \cos \left( \frac{n \pi z}{ L } \right) \vec{\hat{z}}  \, ,
\end{align}
where $L$ is the thickness of the target chip, $n \geq 1$ indexes the modes, and we have focused on the $\vec{\hat{z}}$ direction to enhance coupling to the qubit. A more detailed discussion of all phonon modes is given in Sec.~\ref{subsec:PhononModes}, which includes the full $x,y$-dependence of Eq.~\eqref{eq:simple_mode_function}. The frequency of each phonon is approximately,
\begin{align}
    \frac{\omega_n}{2\pi} = \frac{n c_\text{l}}{2L} \sim 4 \, \text{GHz} \, \left( \frac{n}{300} \right) \left( \frac{c_\text{l}}{10 \, \text{km}/\text{s}}  \right) \left( \frac{400 \, \mu \text{m}}{L} \right)
    \label{eq:omega_n}
\end{align}
where $c_\text{l}$ is the longitudinal sound speed in the target chip. We see that GHz frequency phonons necessitate large $n$ in Eq.~\eqref{eq:omega_n}, which is why the $h$BAR is referred to as a ``high-overtone" bulk acoustic resonator. The frequency difference between the $n$ and $n + 1$ phonon modes, otherwise known as the free spectral range (FSR), is $\Delta \omega_\text{FSR} / 2 \pi \approx c_\text{l} / 2 L \sim 13 \, \text{MHz} \, \left( 400 \, \mu\text{m} / L \right)$. As mentioned above, these modes are confined radially below the PT and for each $n$ there is a finite set of such modes with different radial profiles. These profiles are shown in Fig.~\ref{fig:phonon_mode_illustration} and discussed further in Sec.~\ref{subsec:PhononModes}. Common materials for the target chip hosting the $h$BAR modes are sapphire (\ce{Al2O3}), SiC, and diamond, chosen in large part for their low intrinsic phonon losses~\cite{GokhaleHBARs,KurosuHBARs}.

The third term in Eq.~\eqref{eq:FullHamiltonian} arises from the PT interacting with the electric field of the qubit, thereby coupling the $h$BAR phonon and qubit modes. This coupling allows an $h$BAR phonon to be swapped into a qubit excitation, which can then be read out. An $\mathcal{O}(1)$ swap efficiency, $\varepsilon_\text{s}$, and small, $\mathcal{O}(\mu\text{s})$, swap time, $T_\text{s}$, can be achieved if the qubit frequency is tuned to the phonon mode of interest, $\omega_\text{q} \approx \omega_n$, and $g / 2 \pi \sim \mathcal{O}(100 \, \text{kHz})$. Specific values for $g$, $\varepsilon_\text{s}$, and $T_\text{s}$, are shown in Fig.~\ref{fig:swap_efficiency} and derived in Sec.~\ref{subsec:phonon_qubit_swap_efficiency}. Larger couplings are generally more desirable as they reduce the swap time, $T_\text{s} \sim \pi / 2 g$ (Eq.~\eqref{eq:analytic_estimate}), and in practice can be achieved by choosing materials with a strong piezoelectric response and by maximizing the qubit electric field present in the $h$BAR (e.g., by minimizing the gap between the target and readout chips). However if $g / 2 \pi$ \(\gtrsim\Delta \omega_\text{FSR}\), (corresponding to very wide avoided level crossings in Fig.~\ref{fig:DeviceDesignDiagram}), the qubit may couple quasi-efficiently to multiple phonon modes, which is detrimental to the swap efficiency. A well-tuned \(g\) is therefore desired. Common materials for the PT are AlN, GaN, \ce{LiNbO3}, and NbN, selected for their strong piezoelectric response and good acoustic matching to the substrate materials~\cite{Chu_2017,GokhaleHBARs,KurosuHBARs}. In this work we use a piezoelectric coefficient for AlN of \(e_{\text{pt}} = 0.39 \, \text{N} / \text{m} / \text{V} \) from Ref.~\cite{Chu_2017}, but note that first principles calculations suggest that this may be a conservative value~\cite{deJongAlNPiezo}.

\subsection{Phonon Modes}
\label{subsec:PhononModes}

\begin{figure*}
    \centering
    \includegraphics[width=0.8\linewidth]{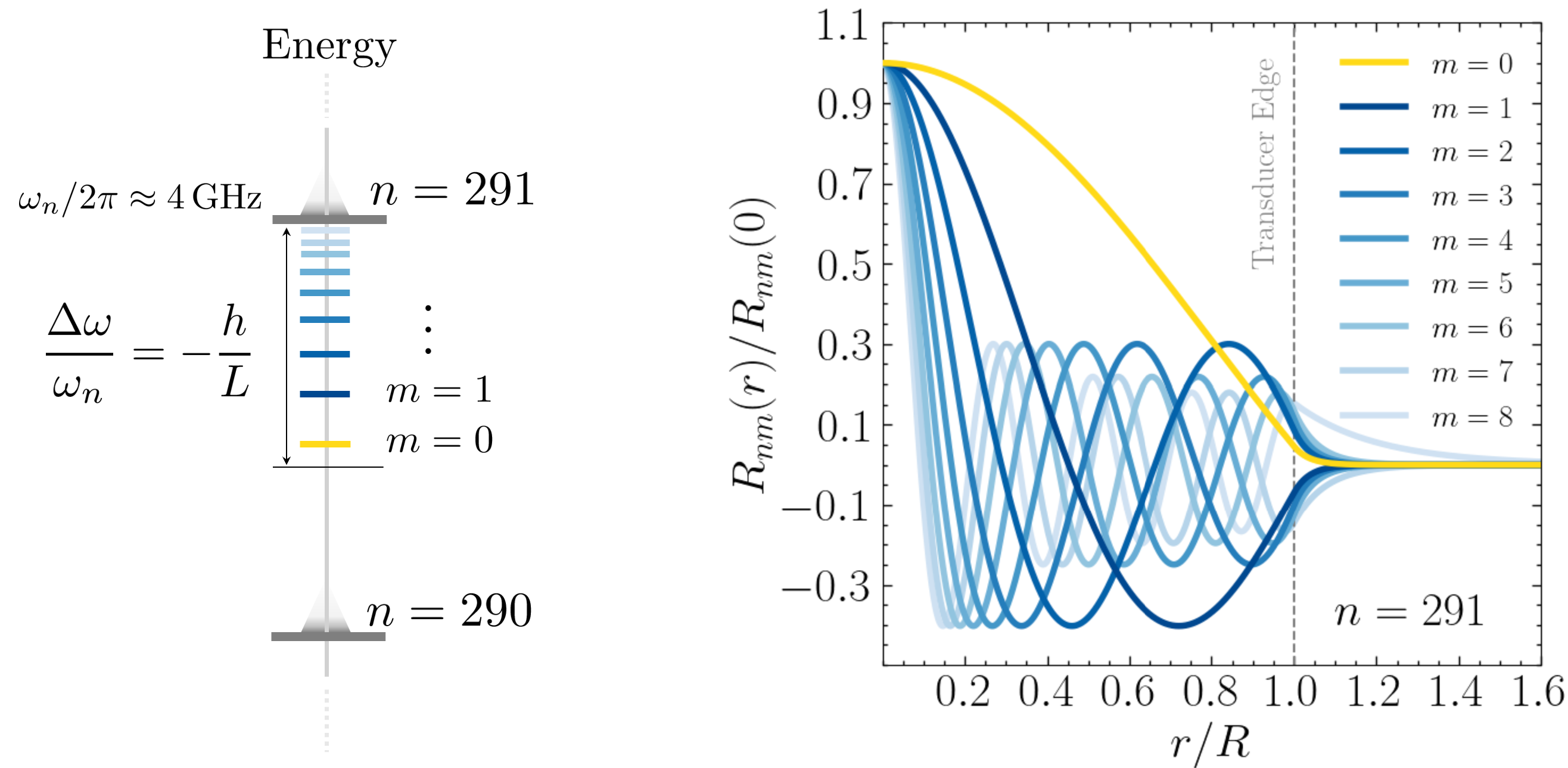}
    \caption{Overview of the phonon modes in the $h$BAR assuming the prototype device parameters in Table~\ref{tab:experimental_parameters}. \textbf{Left}: Schematic of the energy levels in the $h$BAR near $\omega / 2\pi \approx 4 \, \text{GHz}$ ($n = 291$). At each $n$ there are a finite number ($N_{n}$, Eq.~\eqref{eq:N_nm}) of localized modes with energies negatively detuned from $\omega_n = n \pi c_\text{l} / L$. For this prototype device at $n=291$, $N_{n}=9$. The maximum possible detuning is $|\Delta \omega| = \omega_n h / L$. Modes with positive detuning extend laterally throughout the entire chip and are discussed in App.~\ref{app:hbar_eigenmodes}. The color of the energy level of each $m$ mode corresponds to its profile in the right panel. \textbf{Right}: Normalized radial profile of each $m$ phonon mode. The edge of the PT at $r = R$ is labeled ``Transducer Edge".}
    \label{fig:phonon_mode_illustration}
\end{figure*}

There is a diverse assortment of phonon modes in the $h$BAR. Without the PT, the $h$BAR would host bulk phonon modes which extend throughout the entire chip. However, introducing the PT changes the boundary conditions, allowing for ``bound" phonon modes that are spatially confined below the PT in addition to modes extending throughout the entire chip. A detailed derivation of the phonon eigensystem can be found in App.~\ref{app:hbar_eigenmodes}. As discussed above, our focus here will be on the ``bound" modes since they couple most strongly to the qubit. See Refs.~\cite{10.1121/1.1912835,Goryachev:2014yra,hBAR_masters} for similar discussions of the phonon eigensystem in other acoustic resonator devices.

The $h$BAR phonons with the largest coupling to the qubit fluctuate primarily in the $\hat{\vec{z}}$ direction~\cite{Chu_2017}, allowing the displacement vector to be approximated as $\vec{u} \approx u \, \hat{\vec{z}}$. The time-independent displacement operator, $u$, can be quantized in terms of the phonon modes as,
\begin{align}
    & u(r, z) \approx \sum_{n m} \sqrt{\frac{1}{2 \pi \rho_h \omega_{nm} L}} \, U_{nm}(r, z) \left( b_{nm}^\dagger + b_{nm} \right) \\ 
    & U_{nm}(r, z) = \cos{\left( \frac{n \pi z}{L + h(r)} \right)} R_{nm}(r) \, ,
    \label{eq:displacement}
\end{align}
where $\rho_h$ is the $h$BAR mass density, $\omega_{nm}$ is the phonon mode energy, $h(r) = h \, \Theta(R - r)$ is the height profile of the PT, $R$ is the radius of the PT, and the mode functions $R_{nm}(r)$ are unit-normalized such that $\int R^2_{nm} \, r \, dr = 1$. Note that we have assumed there is good acoustic matching between substrate and PT, i.e., they have similar mass densities and sound speeds. See Table~\ref{tab:experimental_parameters} for the parameters specific to the experimental design discussed in Sec.~\ref{sec:DeviceDesignAndOperation}.

The mode numbers, $n, m$, correspond to fluctuations in the $\hat{\vec{z}}, \hat{\vec{r}}$ directions, respectively.\footnote{There are additional modes which oscillate in $\hat{\vec{\phi}}$. However, since the qubit field $\vec{E}_\text{q}$ (Eq.~\eqref{eq:qubit_electric_field}) is approximately constant in $\phi$, only the phonon modes uniform in $\phi$ couple to the qubit.} The radial mode functions, $R_{nm}$, are given by, 
\begin{align}
    R_{nm}(r) & = \frac{\mathcal{A}_{nm}}{R} \begin{cases}
        \displaystyle J_0\left( \frac{\alpha_{nm} \, r}{R} \right) & r \leq R \\[2ex]
        \displaystyle \frac{J_0(\alpha_{nm})}{K_0(\beta_{nm})} \, K_0\left( \frac{\beta_{nm} \, r}{R} \right) & r > R \, ,
    \end{cases} \label{eq:radial}
\end{align}
where $J_i, K_i$ are Bessel functions of the first and second kind, respectively, and,
\begin{align}
    \mathcal{A}_{nm}^2 & = \frac{2 K_0^2(\beta_{nm})}{J_0^2(\alpha_{nm}) K_1^2(\beta_{nm}) + J_1^2(\alpha_{nm}) K_0^2(\beta_{nm})} \\[1ex] 
    \beta_{nm}^2 & = \chi_n^2 - \alpha_{nm}^2~~~,~~~\chi_n = \frac{\omega_n R}{c_\text{t}} \sqrt{ \frac{2 h}{L} } \, . \label{eq:beta}
\end{align}
$\mathcal{A}_{nm}$ is a dimensionless normalization coefficient, $c_\text{t}$ is the transverse speed of sound, and $\alpha_{nm}$ must be solved for numerically by requiring both $0 < \alpha_{nm} < \chi_n$ and continuity of $d R_{nm} / d r$ at $r = R$. An illustration of these modes is given in Fig.~\ref{fig:phonon_mode_illustration}.

For each $n$ there are only a finite number of solutions, $N_{n}$. As discussed in App.~\ref{app:hbar_eigenmodes}, the $\alpha_{nm}$ solutions appear at approximately $\alpha_{nm} \approx \pi m$, which combined with the requirement that $\alpha_{nm} < \chi_n$ gives an estimate for number of solutions at each $n$,
\begin{align}
    N_{n} \sim 9 \, \left( \frac{\omega_n/2 \pi}{4 \,\text{GHz} } \right) \left( \frac{R}{100 \, \mu \text{m}} \right) \left( \frac{h/L}{2.5 \times 10^{-3}} \right)^{1/2} \, ,
    \label{eq:N_nm}
\end{align}
where $c_\text{t}$ is taken from Table~\ref{tab:experimental_parameters}. The energy of each phonon mode has two contributions: the dominant contribution from the oscillations in the $\hat{\vec{z}}$ direction, $\omega_n$ in Eq.~\eqref{eq:omega_n}, and a small $m$ dependent ``detuning", $\delta_{nm}$, from oscillations in the $\hat{\vec{r}}$ direction,
\begin{align}
    \omega_{nm} & = \omega_n - \delta_{nm}~~~,~~~\delta_{nm} = \frac{\omega_n}{2} \left( \frac{c_\text{t} \beta_{nm}}{\omega_n R} \right)^2 
    \label{eq:detuning}
\end{align}
where $\delta_{nm} \ll \omega_n$ for all $m$. The detuning of these bound modes is negative because $\omega_n$ is the frequency of the $n^\text{th}$ mode for a box of length $L$; if we instead defined $\omega_n$ with respect to the total length $L + h$, $\omega_n' \equiv n \pi c_\text{l} / (L+h) \approx \omega_n - \omega_n h / L$, the detuning from $\omega'_n$ would be positive. Furthermore, since $\alpha_{nm} > 0$, the maximum detuning occurs when $\beta_{nm} = \chi_n$. Substituting $\chi_n$ into Eq.~\eqref{eq:detuning}, the maximum detuning is then simply $\delta_{nm}^\text{max} = \omega_n \, h / L$, which can be understood parametrically as,
\begin{align}
    \frac{\delta_{nm}^\text{max}}{2 \pi} \sim 10 \, \text{MHz} \left( \frac{\omega_n / 2 \pi}{4 \,\text{GHz} } \right) \left( \frac{h/L}{2.5 \times 10^{-3}} \right) \, .
\end{align}
Since the maximally detuned mode has the lowest energy in the spectrum at each $n$, we label it with $m = 0$.

To summarize, for each $n$ there are $N_{n}$ bound phonon modes localized near the PT (Eq.~\eqref{eq:N_nm}), with radial mode functions given by Eq.~\eqref{eq:radial} and shown in Fig.~\ref{fig:phonon_mode_illustration}. The energy of each mode is given by Eq.~\eqref{eq:detuning}, and the lowest energy, $m = 0$, mode is maximally detuned by $\omega_n \, h / L$. From here on, we use $\omega_{nm}$ to denote phonon frequencies unless otherwise noted.

\renewcommand{\arraystretch}{1.4}
\setlength{\tabcolsep}{12pt}
\begin{table*}[ht!]
    \begin{center}
        \begin{tabular}{llcccc} \toprule
        & & & \multicolumn{3}{c}{Development Stage} \\\toprule
        & & & \color{Dandelion}{\textbf{Prototype}} & \color{BurntOrange}{\textbf{Gen.} I} & \color{BrickRed}{\textbf{Gen.} II} \\\cmidrule{4-6}
        \textbf{\hbarName}  & PT Height & $h$ & $0.8 \, \mu\text{m}$ & $-$ & $-$ \\
        & PT Piezoelectric Coeff. & $e_\text{pt}$ & $0.39 \, \text{N}/\text{m}/\text{V}$ & $-$ & $-$\\ 
        & Sound Speeds (T, L) & $(c_\text{t}, c_\text{l})$ & $(6, 11) \, \text{km}/\text{s}$ & $-$ & $-$ \\ 
        & Phonon Quality Factor & $Q_\text{p}$ & $10^6$ & $-$ & $-$ \\
        & PT Radius & $R$ & $100 \, \mu\text{m}$ & $400 \, \mu\text{m}$ & $-$ \\ 
        & Height & $L$ & $400 \, \mu\text{m}$ & $40 \, \mu\text{m}$ & $-$ \\
        & Multiplicity & $N_h$ & $1$ & $-$ & $10^3$ \\\hline
        \textbf{Qubit} & Frequency Range & $(\omega_\text{q}^\text{min}, \,\omega_\text{q}^\text{max}) / 2 \pi$ & $(1, 10) \, \text{GHz}$ & $-$ & $-$ \\ 
        & Electric Field at PT & $E_\text{q}$ & $2.9 \times 10^{-2} \, \text{V} / \text{m}$ & $-$ & $-$ \\
        & Qubit Quality Factor & $Q_\text{q}$ & $2.8\times10^{5}$ & $2.8\times10^{7}$ & $-$ \\\hline
        \textbf{Background} & Qubit Energy Gap & $\Delta$ & $176\, \mu\text{eV}$ & $-$ & $-$ \\
        & Normalized QP Density & $x_\text{qp}$ & $10^{-6}$ & $10^{-9}$ & $-$ \\
        & $h$BAR Temp. & $T_{h}$ & $20 \, \text{mK}$ & $5 \, \text{mK}$ & $-$ \\
        & Effective Qubit Temp. & $T_\text{q}^\text{eff}$ & $60 \, \text{mK}$ & $20 \, \text{mK}$ & $-$ \\ 
        & Single Shot Fidelity & $\mathcal{F}$ & $1 - 5 \times 10^{-3}$ & $1 - 5 \times 10^{-4}$ & $-$ \\
        & Sys. $p_\text{dc}$ Fluctuation & $\delta p_\text{dc} / p_\text{dc}$ & $10^{-1}$ & $10^{-4}$ &  $-$.  
        \\\bottomrule
        \end{tabular}
    \end{center}
    \caption{Experimental parameters for the qubit-coupled $h$BAR (qc-$h$BAR) detector at different stages of development (``Prototype", ``Gen. I", and ``Gen. II") assuming an $h$BAR composed of an \ce{Al2O3} chip and \ce{AlN} PT, and an Al qubit. Entries with dashes assume the same value as in the preceding column. Generally, parameters in the Prototype column have been demonstrated today, those for Gen. I are projections for a device optimized for a new physics search, and Gen. II is composed of $N_{h}$ Gen. I detectors operating simultaneously. Detailed discussion of the background parameters can be found in Sec.~\ref{subsec:Backgrounds}.}
    \label{tab:experimental_parameters}
\end{table*}

\subsection{Phonon-Sensing Readout Scheme}
\label{subsec:PhononSensingReadoutScheme}

We now present a scheme to read out the single phonons excited in the $h$BAR. The fundamental operation is the measurement of a qubit excitation created from swapping an $h$BAR phonon with energy $\omega_{nm} = \omega_\text{q}$ into the qubit. This readout scheme, illustrated in Fig.~\ref{fig:DeviceDesignDiagram}, is split into three steps: a search period, an $h$BAR-to-qubit swap period, and a qubit readout period. 

Over the search period, phonons may be excited by a new physics signal. The search period is selected to be the phonon lifetime, $\tau_{nm}$, which we parameterize in terms of a mode-independent phonon quality factor, $Q_\text{p}$, as $\tau_{nm}\equiv Q_{\text{p}}/\omega_{nm}$. After $\tau_{nm}$, the excitation probability saturates as the excitation rate competes with phonon decay. During this period, the qubit is kept detuned from the $h$BAR phonon to restrict swapping. At the end of the search period, the qubit undergoes an active reset, in which it is set into its ground state to ensure that a successful subsequent swap excites the qubit.

In the swap period, the qubit is tuned to resonate with an $h$BAR phonon, $\omega_\text{q} = \omega_{nm}$, which may be done either via flux tuning through a SQUID loop, as shown in Fig.~\ref{fig:DeviceDesignDiagram}, or through a Stark shift as done in Ref.~\cite{Chu_2017}. The qubit is then held on resonance for one Rabi swap time, $T_\text{s}$, which maximizes the probability to convert an $h$BAR phonon to a qubit excitation. The optimal value of $T_\text{s}$ for different phonon modes and device parameters is shown in Fig.~\ref{fig:swap_efficiency}. 

Once the swap time has elapsed, the qubit is again detuned from the $h$BAR phonon to end the swap, and the readout period begins. During the readout period, any qubit excitation may be dispersively read out through the attached resonator~\cite{BlaisCQED,SchusterThesis}. The result of this measurement scheme is a 0 or 1, with 1 (0) representing the excited (ground) state of the qubit and a potential detection (non-detection) of an $h$BAR phonon.

The measurement scheme described above is a minimal set of steps to perform a search for new physics. More sophisticated readout sequences can be used to mitigate dark count backgrounds (Sec.~\ref{subsec:Backgrounds}) attributable to the qubit or $h$BAR. For example, a ``differential'' measurement scheme may be used, in which the above sequence is run once to search for a phonon at a \hbarName{} mode \(n=n_1,\ m=0\) and then run again immediately afterward to search for a phonon in a ``sideband'' mode, \(n=n_1+1,\ m=0\) (with the two-measurement set repeated indefinitely). In searches for small-linewidth new physics that affects only a single phonon mode, this scheme allows for identification of slowly-varying backgrounds common to both modes, such as those from the readout qubit. 

Other sideband measurements may also be useful. For example, a sideband measurement where the qubit is kept off an $h$BAR phonon resonance will provide information on qubit-specific dark count backgrounds. Additional sideband measurements over a broader set of frequencies may help identify frequency dependence of backgrounds in the qubit and $h$BAR. However, in the sections that follow we focus on the simple differential scheme described in the previous paragraph.

\section{Device Performance}
\label{sec:DevicePerformance}

We now focus on parameterizing the qc-$h$BAR performance and calculating the signal strength required for detection in a generic qc-$h$BAR. In Sec.~\ref{subsec:phonon_qubit_swap_efficiency} we determine the efficiency for an $h$BAR phonon to be swapped for a qubit excitation (Fig.~\ref{fig:swap_efficiency}). In Sec.~\ref{subsec:Backgrounds} we estimate the backgrounds for the design detailed in Sec.~\ref{sec:DeviceDesignAndOperation} and comment how these can be reduced in the future. Finally, in Sec.~\ref{subsec:signal_significance} we derive a threshold value for the signal probability in a given measurement step, which must be overcome to claim detection. 

Table~\ref{tab:experimental_parameters} contains a list of the specific parameters assumed and is split into three development stages: ``Prototype", ``Gen. I", and ``Gen. II". ``Prototype" parameters have been demonstrated with technology today, ``Gen. I" parameters assume a geometry more optimized for a new physics search, as well as projections for improvements in backgrounds. ``Gen. II" is a scaled-up version of ``Gen. I", achievable using multiple qubits on a single readout chip via multiplexing (in the case of a 2D transmon design) or multiple 3D cavities operated concurrently in a fridge (in the case of a 3D transmon design). 

\subsection{Phonon-Qubit Swap Efficiency}
\label{subsec:phonon_qubit_swap_efficiency}

\begin{figure*}[ht!]
    \centering
    \includegraphics[width=\textwidth]{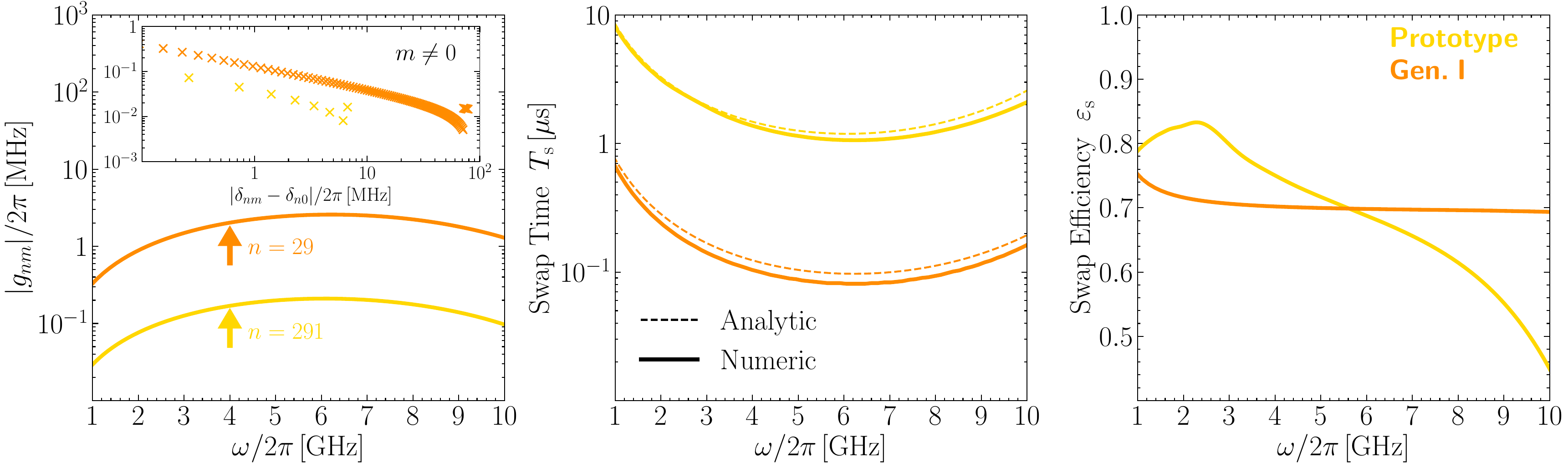}
    \caption{Overview of the parameters governing the $h$BAR phonon-qubit swap efficiency and swap time discussed in Sec.~\ref{subsec:phonon_qubit_swap_efficiency} assuming a qubit frequency $\omega$. Yellow and orange lines assume the ``Prototype" and ``Gen. I" parameters in Table~\ref{tab:experimental_parameters}, respectively. The lines are shown as continuous since the mode separation, $\sim 14 \, \text{MHz} \, (400 \, \mu\text{m} / L)$ is small. \textbf{Left}: Magnitude of the $h$BAR phonon-qubit coupling, $g_{nm}$ (Eq.~\eqref{eq:couplings}), for the $m = 0$ mode. The inset shows the coupling dependence for the $m \neq 0$ mode at $\omega / 2 \pi \approx 4 \, \text{GHz}$, corresponding to $n = 291$ and $n = 29$ modes for the ``Prototype" and ``Gen. I" development stages, respectively. \textbf{Middle}: Swap time, $T_\text{s}$, of the $m = 0$ modes. The dashed line is an analytic estimate, $T_\text{s} \approx \pi / (2 |g_{n 0}|)$, discussed near Eq.~\eqref{eq:analytic_estimate}. \textbf{Right}: Swap efficiency, $\varepsilon_\text{s}$, of the $m = 0$ modes. Note that while the swap efficiency is degraded at small frequencies in Gen. I relative to the Prototype, the overall device sensitivity to new physics is generally improved with larger couplings, and lower swap times can reduce qubit dependent backgrounds.}
    \label{fig:swap_efficiency}
\end{figure*}

The $h$BAR phonon-qubit swap efficiency is the probability of a qubit excitation given a phonon excitation in the $h$BAR. The swap efficiency is generally oscillatory in time once the qubit is brought into resonance with the $h$BAR. To select an on-resonance duration that maximizes the swap efficiency, it is therefore crucial to understand the dynamics of these Rabi oscillations in the $h$BAR phonon-qubit system. The qubit and $h$BAR phonon systems have been discussed in detail in Secs.~\ref{sec:DeviceDesignAndOperation} and~\ref{subsec:PhononModes}, respectively, so we focus on deriving their interaction here.

This qubit will couple to the PT via the electric field it generates, $\vec{E}_\text{q}$. Assuming the PT is small and located near the qubit, as in the setup discussed in Sec.~\ref{sec:DeviceDesignAndOperation}, we can approximate $\vec{E}_\text{q}$ in the PT as,
\begin{align}
    \vec{E}_\text{q} \approx E_\text{q} \, (a + a^\dagger) \, \hat{\vec{z}} \, .
    \label{eq:qubit_electric_field}
\end{align}
Furthermore, though our approach is mostly agnostic to the specific qubit design, we assume that a value of the electric field magnitude matching that of the setup in Ref.~\cite{Chu_2017} is reasonable, and set $E_\text{q} \approx 2.9 \times 10^{-2} \, \text{V} / \text{m}$.

The qubit electric field in Eq.~\eqref{eq:qubit_electric_field} couples to the polarization of the PT, $\vec{P}$, via $H_\text{int} = - \int_\text{pt} d^3\vec{x} \, \vec{P} \cdot \vec{E}_\text{q}$, where the integral is over the volume of the PT. The polarization of the PT is $\vec{P}^i = e^{ijk}_\text{pt} \nabla^j \vec{u}^k$, where $e^{ijk}_\text{pt}$ is the piezoelectric coefficient of the PT. Focusing on the $\hat{\vec{z}}$ direction simplifies the interaction Hamiltonian to,
\begin{align}
    H_\text{int} & \supset \sum_{nm} g_{nm} \left( a^\dagger \, b_{nm} + b_{nm}^\dagger \, a \right) \label{eq:h_int} \\
    g_{nm} & = e_\text{pt} E_\text{q} \, \sqrt{ \frac{2 \pi R^2}{ \rho_h \, \omega_{nm} \, L} } \, \mathcal{B}_{nm} \label{eq:couplings} \\ 
    \mathcal{B}_{nm} & = - \mathcal{A}_{nm} \left[ (-1)^n - \cos{\left( \frac{n \pi L}{L + h} \right)} \right] \, \frac{J_1(\alpha_{nm})}{ \alpha_{nm} } \, ,
    \label{eq:B_nm}
\end{align}
where we have kept only the energy-conserving terms, $e_\text{pt} \equiv e_\text{pt}^{zzz}$, and $\mathcal{B}_{nm}$ is a dimensionless form factor. The numerical values for $g_{nm}$ given the parameters in Table~\ref{tab:experimental_parameters} are shown in the left panel of Fig.~\ref{fig:swap_efficiency}. The solid line shows the coupling of the most bound ($m=0$) mode, with the inset showing that high $m$ modes have significantly lower coupling.

From Eq.~\eqref{eq:couplings} we see that a larger $g_{nm}$ can be achieved in a variety of ways. First the geometrical shape can be optimized, with a larger PT radius, $R$, and smaller $h$BAR thickness $L$ being beneficial. For the prototype development stage we assume $R, L$ values similar to Ref.~\cite{Chu_2017}, and for Gen. I optimize them further. The maximum $R$ is set by the lateral size of the electric field in the PT, which is governed by the lateral dimensions of the qubit. For the Gen. I development stage, we assume that a transmon qubit can be designed to efficiently couple to a PT with radius $R=400~\mu$m. However, larger qubits have larger capacitance and correspondingly lower anharmonicities. This anharmonicity must be kept sufficiently large to prevent spurious \(\ket{e}\rightarrow\ket{f}\) transitions during fast control gates such as the swap. The minimal $L$ is set by fabrication constraints, but should be much larger than $h$ for our approximation of the system to hold. Smaller $L$ can also be beneficial for the detector operation since it increases the FSR. The PT height, $h$, is also chosen to optimize the couplings. Note that the bracketed term in $\mathcal{B}_{nm}$ (Eq.~\eqref{eq:B_nm}) is approximately maximized when $n$ is an odd integer multiple of $L / h$ when $h \ll L$. Therefore for a given $h$ the difference in frequency between peaks of $g_{nm}$ is $\Delta \omega / 2 \pi = c_\text{l} / h$ (see Eq.~\eqref{eq:omega_n}). $h$ should therefore be chosen such that $\Delta \omega / 2 \pi$ is a bit larger than the frequency range of interest, $(\omega^\text{max} - \omega^\text{min}) / 2 \pi$, and such that $n$ is an odd integer multiple of $L/h$ near the center of the frequency range. We find $h = 0.8 \, \mu\text{m}$ sufficient. Additionally, the material dependent parameters, e.g., $e_\text{pt}$, can also be optimized. However in order to approximate the target chip as a single material the $h$BAR and PT should have similar sound speeds and densities, which motivates careful materials selection as discussed in Ref.~\cite{GokhaleHBARs}.

Given the interaction Hamiltonian in Eq.~\eqref{eq:h_int} we can derive the $h$BAR phonon-qubit evolution equation at each $n$, for all $N_n$ (Eq.~\eqref{eq:N_nm}) bound modes. Decomposing the $h$BAR phonon-qubit state as a linear combination, $| \psi(t) \rangle = c_\text{q}(t) | e \rangle + \sum_{m} c_{nm}(t) | n m \rangle$, the evolution equation can be written in matrix form as,
\begin{align}
    \begin{bmatrix}
        \dot{c}_\text{q} \\ \dot{c}_{n0} \\ \dot{c}_{n1} \\ \vdots 
    \end{bmatrix} & = \mathbf{M} \begin{bmatrix}
        c_\text{q} \\ c_{n0} \\ c_{n1} \\ \vdots 
    \end{bmatrix} \label{eq:evolution_equation} \\ 
    \mathbf{M} = &\begin{bmatrix}
        - \gamma_\text{q} + i \,\omega_\text{q} & i\, g_{n0} & i \,g_{n1} & \cdots \\
        i \,g_{n0} & - \gamma_{n0} + i \, \omega_{n0} & & \\ 
        i \,g_{n1} & & - \gamma_{n1} + i \, \omega_{n1} & \\ 
        \vdots & & & \ddots \\
    \end{bmatrix} \, ,
\end{align}
where $\gamma_{nm} = \omega_{nm} / Q_\text{p}$ and $\gamma_\text{q} = \omega_\text{q} / Q_\text{q}$ are the phonon and qubit linewidths, respectively. Solving the system of equations in Eq.~\eqref{eq:evolution_equation} gives the probability of a qubit excitation at any time, $|c_\text{q}(t)|^2$. The swap time, $T_\text{s}$ is defined as the time at which $|c_\text{q}(t)|^2$ is maximized, and the swap efficiency, $\varepsilon_\text{s} \equiv |c_\text{q}(T_\text{s})|^2$, is the maximal swap probability.

In Fig.~\ref{fig:swap_efficiency} we show the results of solving Eq.~\eqref{eq:evolution_equation}, assuming the device parameters given in Table~\ref{tab:experimental_parameters}, for $1 \, \text{GHz} < \omega_\text{q} / 2 \pi < 10 \, \text{GHz}$. This frequency range corresponds to the range of mode numbers, $73 \, (L / 400 \mu\text{m})~\lesssim~n~\lesssim~730 \, (L / 400 \mu\text{m})$. The solid lines are found by matching the qubit frequency, $\omega_\text{q}$, to the frequency of the most bound phonon mode, $\omega_{n0}$, for each $n$ and then solving Eq.~\eqref{eq:evolution_equation} assuming an initial condition $c_{n0} = 1$, with all other $c$'s zero. This maximizes the swap efficiency at each $n$ since the $m = 0$ phonon modes have the largest coupling to the qubit. Lastly we note that the swap time, $T_\text{s}$, shown in the middle in Fig.~\ref{fig:swap_efficiency}, can be understood analytically as
\begin{align}
    T_\text{s} & \approx \frac{\pi}{2 \, |g_{n0}|} \, ,
    \label{eq:analytic_estimate}
\end{align}
which can be derived analytically from a simple two-state model with coupling $g_{n 0}$ between an $h$BAR phonon and qubit. These analytic solutions are shown in Fig.~\ref{fig:swap_efficiency} as dashed lines, and match the full numerical results reasonably well. While a two-state model captures the swap time well, the presence of the $m \neq 0$ modes still impacts the swap efficiency. The right panel of Fig.~\ref{fig:swap_efficiency} shows that swap efficiency is somewhat reduced with ``Gen. I'' parameters, despite the larger coupling and shorter swap time. This is due to the increased number of bound modes with $m \neq 0$, as can be seen in the inset of the left panel. There is a nonzero probability to be found in these additional bound modes, which slightly reduces the overall probability that the excitation ends up in the qubit after one swap time.

\subsection{Backgrounds}
\label{subsec:Backgrounds}

\begin{figure*}
    \centering
    \includegraphics[width=\linewidth]{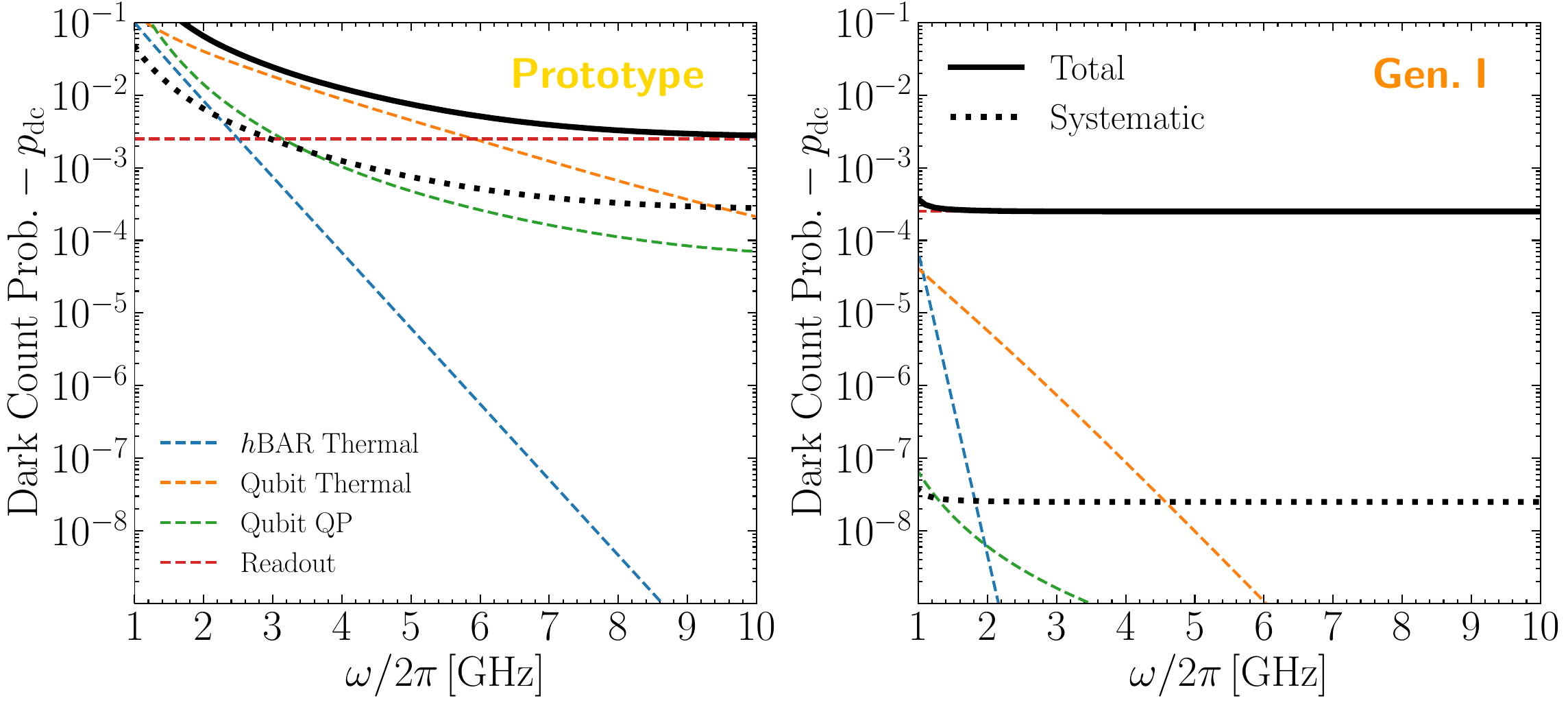}
    \caption{Dark count probability at each measurement step, $p_\text{dc}$ (Eq.~\eqref{eq:p_dc}), for the Prototype (left) and Gen. I (right) parameters from Table~\ref{tab:experimental_parameters} is shown in black, assuming an operating frequency of $\omega / 2 \pi$. Each of the components of $p_\text{dc}$, discussed in detail in Sec.~\ref{subsec:Backgrounds}, are also shown as dashed lines. ``$h$BAR Thermal" (blue) corresponds to thermal fluctuations in the $h$BAR, ``Qubit Thermal" (orange) corresponds to effective thermal fluctuations in the qubit, ``Qubit QP" (green) corresponds to quasiparticle excitations of the qubit,  and ``Readout" (red) corresponds to readout mis-measurement error. The black lines, illustrating $p_\text{dc}$, are the sum of the dashed colored lines. Additionally we show a black dotted line corresponding to the assumed level of systematic fluctuations, $\delta p_\text{dc}$.}
    \label{fig:frequency_dependence_of_pdc}
\end{figure*}

We now consider sources of ``dark-count"-like backgrounds which can cause the measurement to return ``1'' in the absence of a new physics signal. The probability of a dark count appearing in any measurement step is defined as $p_\text{dc}$. To first order, i.e., for small dark count probabilities, $p_\text{dc}$ can be decomposed into several components,
\begin{equation}
    \label{eq:p_dc}
    p_{\text{dc}} = p_{\text{th},h} + p_{\text{th,q}} + p_{\text{qp}} + p_{\text{ro}} \, .
\end{equation}
$p_{\text{th},h}, p_\text{th,q}$ are from thermal excitations of the $h$BAR and qubit, respectively, $p_\text{qp}$ are due to quasiparticle-induced excitations of the qubit, and $p_\text{ro}$ is due to errors in the readout step. Each of these contributions to $p_\text{dc}$ is shown in Fig.~\ref{fig:frequency_dependence_of_pdc} for the parameters given in Table~\ref{tab:experimental_parameters}. Estimates for the prototype and Gen. I development stages broadly represent the backgrounds achievable today and projections for the future based on recent development, respectively. We now discuss each of these background sources in detail below. Additional details can be found in App.~\ref{app:backgrounds}.

\vspace{1mm}
\noindent
\textbf{Thermal $h$BAR Phonons} \( ( p_{\text{th},h} ) \): For an $h$BAR in thermal equilibrium at temperature $T_{h}$, the probability of a thermal excitation in a mode with energy $\omega$ is $p_{\text{th},h} = ( e^{ \omega / T_{h} } - 1 )^{-1}$. The time to reach thermal equilibrium is set by the $h$BAR phonon lifetime, $\tau_{nm} = Q_\text{p} / \omega_{nm}$. Since our measurements are separated by at least a phonon lifetime it is fair to assume the $h$BAR mode of interest has reached thermal equilibrium. Given the range of achievable base temperatures in typical dilution fridges, the prototype and Gen. I estimates assume $T_{h} = 20 \, \text{mK} $ and $T_{h} = 5 \, \text{mK}$, respectively.

\vspace{1mm}
\noindent
\textbf{Thermal Qubit Excitation} \( ( p_{\text{th,q}} ) \): The qubit can also be excited from a surrounding thermal bath with effective temperature \(T_{\text{q}}^{\text{eff}}\)~\cite{Serniak2018}. Such baths may arise from a collection of two-level systems which are known to exchange energy with qubits~\cite{Kulikov,MullerTLSReview,NonMarkovianPhononicBandgap}. For a qubit reset into its $\ket{g}$ state, excitations will leak back in and re-establish a thermal occupation probability over the characteristic time \(\tau_\text{q}=Q_\text{q} / \omega_\text{q}\). This ``rethermalization'' process, which has been known to limit the efficacy of unconditional active reset gates~\cite{MagnardActiveReset}, occurs throughout the active reset, swap, and readout pulses. Since the total duration of these pulses \(\Delta t \ll \tau_\text{q}\), the probability of finding a qubit excitation is therefore reduced from its equilibrium value,
\begin{align}
    \label{eq:p_exc_th}
    p_{\text{th,q}} \approx \Bigg[\frac{\Delta t\, \omega_\text{q}}{2Q_\text{q}} \, \frac{1}{e^{\omega_\text{q}/T_{\text{q}}^{\text{eff}}} - 1}\Bigg]\left[ 1+e^{- T_{m} \omega_{nm}/Q_{\text{p}}}\right]
\end{align}
where \(\Delta t \equiv T_{\text{ar}} + T_\text{s} + T_{\text{ro}}\) is the sum of active reset, swap, and readout pulse times. \(T_{\text{ar}}=T_{\text{ro}}=1~\mu s\) is assumed for simplicity, and \(T_\text{m} \approx \tau_{nm}\) is the time between the swap gates of two subsequent measurements at this $h$BAR mode. The first term in Eq.~\eqref{eq:p_exc_th} loosely follows from Ref.~\cite{TobarSingleGraviton}, with the factor of 1/2 coming from the approximation that only excitations in the second half of the reset-swap-readout sequence will be observed as a dark count in this measurement. The second term is a correction that accounts for spontaneous excitations in the first half of the reset-swap-readout period \(\Delta t\), which instead of being directly observed may be efficiently swapped back into the $h$BAR (a detailed diagram of this timeline is given in Fig.~\ref{fig:app_injected_phonons}). While the phonons swapped out don't directly affect the current measurement, if they are swapped back in during the next measurement, at time \(T_\text{m} \approx \tau_{nm}\) later, that is a background.\footnote{Active reset failure more generally can lead to this effect. While we assume active reset failure rates to be limited by qubit spontaneous excitation in the near-term as demonstrated by Ref.~\cite{MagnardActiveReset}, additional engineering may cause such ``injected'' phonon backgrounds to be dominated by other effects. Since these backgrounds can in principle be reduced by simply waiting longer between measurements of a given $h$BAR mode, we keep our near-term assumptions simple and ignore such effects at leading order.} Critically, \(T_{\text{q}}^{\text{eff}}\) is often much higher than the base temperature of a dilution fridge as a result of effects such as poor sample thermalization or microwave noise. As a result, for the prototype setup we assume \(T_{\text{q}}^{\text{eff}}=60\)~mK, as estimated from Ref.~\cite{Serniak2018}, and our Gen. I estimate assumes that we can achieve thermalization down to $T_\text{q}^\text{eff} = 20 \, \text{mK}$, nearly thermalized with the baseplate of a fridge. Furthermore we assume $Q_\text{q} = 2.8 \times 10^5$ ($Q_\text{q} = 2.8 \times 10^7$) which, at $\omega_\text{q} / 2 \pi = 4.5 \, \text{GHz}$, corresponds to $\tau_\text{q} = 10 \, \mu\text{s}$ ($\tau_\text{q} = 1 \, \text{ms}$) for the prototype (Gen. I) stage~\cite{von2022parity}. The Gen. I value is a projection based on the three orders of magnitude gained in qubit lifetime over the last two decades~\cite{SiddiqiReview}, and an assumption that qubit losses due to interactions with uncontrolled degrees of freedom in the PT and $h$BAR can be engineered to be subdominant to losses intrinsic to the qubit. 

\vspace{1mm}
\noindent
\textbf{Quasiparticle-Induced Qubit Excitation} \( ( p_{\text{qp}} ) \):  The qubit can also be spuriously excited during the swap or readout gates via tunneling of nonequilibrium ``hot'' Bogoliubov quasiparticles (QPs) through the qubit's Josephson junction~\cite{Serniak2018,WennerHotQuasiparticles,Jin2015}. Following the set of assumptions in Refs.~\cite{WennerHotQuasiparticles,CatelaniRelaxation}, a semi-quantitative estimate for the probability for this to occur during the swap/readout period is (see App.~\ref{subsubsec:DiscussionOfGeneralBackgroundModel} for more detail):
\begin{align}
    \label{eq:p_exc_qp}
    p_{\text{qp}} \approx&\ \frac{\Delta t}{2}\Bigg[2.17 \Bigg(\frac{\Delta}{\omega_\text{q}}\Bigg)^{3.65} \sqrt{\frac{2\omega_\text{q}\Delta}{\pi^{2}}} x_\text{qp}^{2}\Bigg]\hspace{-0.4em}\left[ 1+e^{- T_{m} \omega_{nm}/Q_{\text{p}}}\right]
\end{align}
where $x_\text{qp}$ is the normalized QP density in the junction leads, $\Delta$ is the qubit's superconducting gap energy, and we again include in the last term the effect of a prior measurement's failed active reset. For our prototype and Gen. I scenarios we assume \(x_\text{qp}=10^{-6}\) and \( x_\text{qp} = 10^{-9}\), respectively~\cite{WangMeasurementAndControl2014,ConnollyYale2024}.

\vspace{1mm}
\noindent
\textbf{Readout Mismeasurement, \( ( p_{\text{ro}} ) \):} Finally, a failed readout of the true qubit ground state will contribute to the dark counts. The ``ideal'' single-shot measurement fidelity (SSF) is a measure of the readout noise, and is defined as \(\mathcal{F} =1-p(1|g)-p(0|e) \approx 1 - 2 \, p(1|g) \), where $p(1|g) = p_{\text{ro}}$ is the probability of measuring a $1$ when the qubit is in $\ket{g}$, and $p(0|e)$ is the probability of measuring $0$ when the qubit is in $\ket{e}$. As the ``ideal'' SSF does not account for the contribution of spontaneous excitations or decays to the probability of mismeasurement, it is approximately symmetric, i.e. $p(1|g) \approx p(0|e)$~\cite{ChenSingleShotFidelity}. Typical experiments routinely achieve \(\mathcal{F} \sim 0.6-0.9 \), but fidelities over 0.995 have been measured, suggesting \(p_{\text{ro}} \simeq 2.5 \times 10^{-3} \)~\cite{WalterSingleShot,ChenSingleShotFidelity}. We take this as our prototype estimate for \(p_{\text{ro}}\), and given the pace of recent developments estimate that within the next ten years this readout error can be improved by over an order of magnitude~\cite{hazraSingleShotFidelity}. For a given qubit and readout resonator hardware configuration, the \(\ket{g}/\ket{e}\) readout separation (and therefore this background) is dependent on the detuning between the qubit and readout resonator, and should nominally vary as a function of \(\omega_\text{q}\). However, since this is dependent on specific details of the fixed readout resonator frequency, and since multiple readout resonators with different frequencies may be used to concurrently read out qubits in different \(\omega_\text{q}\) ranges, we approximate this background to be flat for \(1 \, \text{GHz} \lesssim \omega_\text{q}/2\pi \lesssim 10 \, \text{GHz} \) where commercial microwave hardware is readily available.

The individual background probabilities, $p_{\text{th},h}$, $p_{\text{th,q}}$, $p_{\text{qp}}$, $p_{\text{ro}}$, and their sum, $p_\text{dc}$, are shown as a function of frequency \(\omega=\omega_{nm}=\omega_\text{q}\) in Fig.~\ref{fig:frequency_dependence_of_pdc}. These specific probabilities are presented as estimates to gain an understanding of what the dominant backgrounds will be. Experimental calibration of these backgrounds, e.g., by running our measurement scheme without the swap gate included, is critical and will be an essential component of a first prototype device.

As our search strategy calls for the use of a differential measurement scheme, it is also critical to understand the variation in \(p_{\text{dc}}\), \(\delta p_{\text{dc}}\). In principle, such variation may be modeled using variations in the four components, i.e., $( \delta p_{\text{dc}} )^2 = ( \delta p_{\text{th},h} )^2 + (\delta p_{\text{th,q}})^2 + \cdots$. These variations can be further decomposed in to terms reflecting variations in the model parameters, e.g., $\delta T_{h}, \delta T^\text{eff}_q$ etc. These fluctuations must be taken over the intra-measurement-set time, and as a result, are challenging to estimate {\it{a priori}}. Benchmarking measurements of \(x_\text{qp}\), \(\omega_\text{q}\), and \(\tau_\text{q} \) (\(Q_\text{q}\)) have been made in the context of achieving stable qubit operation for QIS applications~\cite{SchlorFluctuations,BurnettDecoherenceBenchmarking,KlimovFluctuations,Muller2015Fluctuations,NonPoissonianQPsFluxonium}. On timescales of minutes to hours these quantities have been found to experience relative fluctuations at the \(10^{-2}\) to \(5\times10^{-1}\) level. However, due to the statistical nature with which these quantities are defined and measured, such probes have largely been unable to measure variations on much faster timescales than a second, suggesting that significant extrapolation is needed to estimate fluctuations on intra-measurement-set, 10~\(\mu\)s timescales. For our prototype scenario, we make an assumption that \(\delta p_{\text{dc}}/p_{\text{dc}}\sim0.1\) on these short timescales, given the observed behavior in some of the detector parameters on long timescales. Estimates for some of these fluctuations on short timescales may be possible using insights from other superconducting detector communities~\cite{DeVisserNumberFluctuations,WilsonProber}. For our Gen. I we assume $\delta p_\text{dc} / p_\text{dc} \approx 10^{-4}$, reflecting the relative fluctuation needed to be competitive in the physics searches described in Sec.~\ref{sec:DarkMatterSearchModes}. Ultimately, these fluctuations are measurable {\it{in situ}} using the differential readout scheme. 

\subsection{Signal Significance}
\label{subsec:signal_significance}

To search for new physics with the qc-$h$BAR we must be able to identify an excess number of phonons over the expected background. This is equivalent to requiring that the probability of a signal appearing in the search period, $p_\text{s}$, is greater than a threshold value, $p_\text{s}^\text{thr}$, which depends on the dark count probability, $p_\text{dc}$, discussed previously in Sec.~\ref{subsec:Backgrounds}. In this section we estimate $p_\text{s}^\text{thr}$ for the non-differential and differential measurement schemes. These are useful to reduce systematic effects in complementary contexts. The non-differential measurement scheme is simpler and can be used when the signal has specific features, e.g., daily modulation. The differential measurement scheme is useful for subtracting away background contributions common to neighboring $h$BAR phonon modes. While we anticipate more sophisticated statistical methods to be used during a proper data analysis, we discuss these models to highlight the essential elements and offer an understanding of their significance.

We begin with the non-differential measurement scheme. Let us model the result of any individual measurement as an independent Bernoulli random variable, $Q_i$, which in the presence of a signal returns $1$ with probability $\varepsilon_\text{s} p_\text{s} + p_\text{dc}$, and in the absence of a signal returns $1$ with probability $p_\text{dc}$. The random variable describing the result of $N$ measurements is taken to be 
\begin{align}
    \chi_\text{non-diff.} = \frac{1}{N} \sum_{i = 1}^{N} Q_i \, .
\end{align}
To detect new physics $\chi_\text{non-diff.}$ must be much larger in the presence of a signal than in the absence. Mathematically this means that to detect new physics at the $n_\sigma$ significance level we require $\text{E}[\chi_{\text{non-diff.}}]_\text{S+B} > \text{E}[\chi_\text{non-diff.}]_\text{B} + n_\sigma \, \sigma[\chi_\text{non-diff.}]_\text{B} $, where $\text{E}$ is the expectation value and $\sigma$ is the standard deviation. Subscripts represent the assumption under which the expectations are taken: $\text{S+B}$ means signal and background, whereas $\text{B}$ indicates only background. The previous condition defines $p_{\text{s},\text{non-diff.}}^\text{thr}$ at the $n_\sigma$ significance level and is equal to,
\begin{align}
    p_{\text{s},\text{non-diff.}}^\text{thr} \simeq \frac{n_\sigma}{\varepsilon_\text{s}} \sqrt{ \frac{p_\text{dc}}{N}} \, ,
    \label{eq:non_diff}
\end{align}
where we have assumed that all probabilities are small. Note that if there are $N_h$ devices (as in the Gen. II development stage) one simply replaces $N \rightarrow N_h N$. Eq.~\eqref{eq:non_diff} is only the statistical uncertainty, which can be seen by considering how Eq.~\eqref{eq:non_diff} relates to a threshold number of signal events. A detectable signal must have $p_\text{s} > p_{\text{s},\text{non-diff.}}^\text{thr}$, and multiplying both sides of this inequality by $N$ results in $N_\text{s} > N_{\text{s},\text{non-diff.}}^\text{thr}$, where $N_\text{s} \equiv p_\text{s} N$ is the expected number of signal events, $N_{\text{s},\text{non-diff.}}^\text{thr} \equiv p_{\text{s},\text{non-diff.}}^\text{thr} N = (n_\sigma / \varepsilon_\text{s}) \sqrt{ N_\text{dc} }$, and $N_\text{dc} = p_\text{dc} N$ is the expected number of background events. This is the usual criteria for statistical uncertainties, i.e., the signal must be greater than the square root of the background. In addition to statistical uncertainties there may also be systematic uncertainties. For example, if the true dark count probability was slightly larger than that assumed in our background model, the additional background events generated due to this would mimic a signal. Since the value of $p_\text{dc}$ is not well known these systematic biases must be avoided if the signal does not have its own distinguishing features, e.g., if the signal does not modulate daily.

One way to mitigate systematic biases is to use a differential measurement scheme, which we consider now. Each measurement consists of querying two $h$BAR phonon modes: a ``target" mode and a ``sideband" mode (the latter of which is assumed to be insensitive to a new physics signal). Taking $Q_{i,1}$ and $Q_{i,2}$ to be the random variables describing the results of measuring the target and sideband modes, respectively, the search observable is,
\begin{align}
    \label{eq:ChiObservable}
    \chi_\text{diff.} = \frac{1}{N} \sum_{i=1}^{N} \left[ \, Q_{i,1} - Q_{i,2} \, \right] \, ,
\end{align}
which represents the difference in counts between the target and sideband modes. Again, to detect a new physics signal at the $n_\sigma$ significance level we require $\text{E}[\chi_\text{diff.}]_\text{S+B} > \text{E}[\chi_\text{diff.}]_\text{B} + n_\sigma \sigma[ \chi_\text{diff.}]_\text{B}$. In this case the threshold probability is,
\begin{align}
    \label{eq:psmin}
    p_{\text{s},\text{diff.}}^\text{thr} \simeq \frac{n_{\sigma}}{\varepsilon_\text{s}} \left( \sqrt{\frac{2p_{\text{dc}}}{N} } + \delta p_{\text{dc}} \right) \, ,
\end{align}
where the factor of two in the first term, relative to Eq.~\eqref{eq:non_diff}, is due to two $Q_i$'s appearing at each measurement step, and we have explicitly included a contribution from the systematic bias in the dark count probability between modes $1$ and $2$. To understand the definition of $\delta p_\text{dc}$, note that in the limit where the second term in Eq.~\eqref{eq:psmin} is dominant in order for a signal to be significant we must have $p_\text{s} > (n_\sigma / \varepsilon_\text{s}) \delta p_\text{dc}$. Since the significance threshold does not decrease by increasing $N$, $\delta p_\text{dc}$ represents a systematic contribution to the dark count background. Additional systematic uncertainties could also arise due to correlations between the backgrounds at consecutive measurements.\footnote{If the dark count probability does fluctuate then the $p_\text{dc}$ in Eq.~\eqref{eq:psmin} should be replaced with an average value. This can be measured {\it{in situ}} by utilizing other observables. For example, the expectation value of $\lambda = (1/2N) \sum_i [ Q_{i,1} + Q_{i,2} ]$ is the average value of $p_\text{dc}$ between the modes.} As in the non-differential measurement scheme if $N_h$ devices are used then $N \rightarrow N_h N$ in Eq.~\eqref{eq:psmin}.

The differential threshold probability in Eq.~\eqref{eq:psmin} simplifies in ``statistically-limited" and ``systematically-limited" regimes where the first and second terms in Eq.~\eqref{eq:psmin} dominate, respectively. That is,
\begin{align}
    p_{s,\text{diff.}}^\text{thr} = \frac{n_\sigma}{\varepsilon_\text{s}} \begin{cases}
        \displaystyle \sqrt{ 2 \, p_\text{dc} / N }& \text{stat.} \\
        \displaystyle \delta p_\text{dc} & \text{sys.} \, ,
    \end{cases}
    \label{eq:p_s_diff_cases}
\end{align}
where ``stat." and ``sys." indicate the statistically and systematically limited operation regimes, respectively. To find how long an experiment must run to achieve the systematic limit, $T_\text{sys}$, we use the fact that $T_\text{obs} \approx N \tau_{nm} = N Q_\text{p} / \omega$, for operation at frequency $\omega$, and equate the cases in Eq.~\eqref{eq:p_s_diff_cases},
\begin{align}
    T_\text{sys} \sim \frac{\text{yr}}{N_h} \, \left( \frac{10^{-4}}{\delta p_\text{dc} / p_\text{dc}} \right)^2 \left( \frac{10^{-4}}{p_\text{dc}} \right) \left( \frac{Q_\text{p}}{10^6} \right) \left( \frac{4 \, \text{GHz}}{\omega / 2 \pi} \right) \, .
\end{align}

Practical use of Eq.~\eqref{eq:psmin} in a sensitivity or limit-setting capacity clearly requires knowledge of both \(p_{\text{dc}}\) and \(\delta p_{\text{dc}}\). Ultimately both of these, while estimated in detail in Sec.~\ref{subsec:Backgrounds}, will be measured {\it{in situ}}. Initial experiments will be useful for probing the \(N\) at which one deviates from statistically-limited operation, at which point auxiliary studies can be done to understand, and mitigate, fundamental sources of systematic uncertainties. 

\section{Signals}
\label{sec:DarkMatterSearchModes}

There are a variety of different physical effects which can excite phonons in the $h$BAR. If the DM in our galaxy consists of ultralight DM particles which couple to matter, these particles can be absorbed and resonantly excite phonons whose energy, $\omega_{nm}$, is approximately equal to the DM mass, $\omega_{nm} \approx m_\text{DM}$. Additionally, high-frequency gravitational waves (GWs) can also create phonons when the GW frequency is resonant with the $h$BAR phonons, $\omega_{nm} \approx \omega_\text{GW}$. For both signals we consider a regime where the bandwidth is narrow such that only a single phonon mode is excited. This is satisfied for ultralight DM signals and some high-frequency GW signals.

While the frequency of both the ultralight DM and high-frequency GW signals must be resonant with the $h$BAR phonon frequency, the wavelengths of both signals differ dramatically from the phonon wavelength. The incoming wavelengths are $\lambda_\text{DM} = 2 \pi / (\omega \, v_\text{DM}) \sim 75 \, \text{m} \, (2 \pi \cdot 4 \, \text{GHz} / \omega)$ and $\lambda_\text{GW} =  2 \pi / \omega \sim 7.5 \, \text{cm} \, (2 \pi \cdot 4 \, \text{GHz} / \omega)$ for an ultralight DM and high-frequency GW signal, respectively, where $v_\text{DM} \sim 10^{-3}$ is the typical DM velocity. Both are larger than the size of the detector, and much larger than the phonon wavelength, $\lambda_{nm} = 2 \pi c_\text{l} / \omega_{nm} \sim 0.75 \, \mu \text{m} \, (2 \pi \cdot 4  \, \text{GHz} / \omega_{nm})$. This wavelength mismatch has important ramifications when trying optimize the $h$BAR dimensions.

To gain intuition for how the wavelength mismatch affects the $h$BAR phonon excitation rate, consider an interaction Hamiltonian $\delta H(t) = \int d^3 \vec{x} \, \vec{f}(t) \cdot \vec{u} \approx f(t) \int d^3\vec{x} \, u $, which couples a harmonic, long-wavelength force density with frequency $\omega$, $\vec{f}(t) \approx f(t) \, \hat{\vec{z}} \equiv f_0 \, (e^{i \omega t} + e^{- i \omega t})\, \hat{\vec{z}}$, to the $h$BAR phonons. The matrix element squared between the no phonon, $| 0 \rangle$, and single phonon, $| n m \rangle$, state appearing in Fermi's golden rule is then,
\begin{align}
    \left| \langle n m | \delta H_0 | 0 \rangle \right|^2 & = \frac{ 2 \pi f_0^2}{\rho_h \omega_{nm} L} \left( \int U_{nm}(r,z) \, r \,  dr dz \right)^2 \, ,\label{eq:matrix_element_example}
\end{align}
where $\delta H_0 = f_0 \int d^3 \vec{x} \, u$ and is defined via $\delta H(t) \equiv \delta H_0 e^{-i \omega t}~+~\text{h.c.}$. If the phonon mode was uniform in $r, z$ over the $h$BAR, then the integral in Eq.~\eqref{eq:matrix_element_example} would parametrically be of size $\sim R \, L$. However, since the phonon mode oscillates many times in the $\hat{\vec{z}}$ direction, the integral is instead parametrically $\sim R \, \lambda_{nm}$, where $\lambda_{nm} = 2 \pi c_\text{l} / \omega_{nm}$ is the wavelength of the phonon mode. Physically this is because long-wavelength signals only couple coherently to the mass within, roughly, one wavelength of the phonon mode, $M_\lambda \equiv \rho_h \pi R^2 \lambda_{nm}$; contributions from larger length scales average out. We can make these parametrics manifest by rewriting Eq.~\eqref{eq:matrix_element_example} as 
\begin{align}
    \left| \langle n m | \delta H_0 | 0 \rangle \right|^2 & = \left( \frac{f_0}{\rho_h} \right)^2 \frac{2}{\omega_{nm}} \frac{M_\lambda^2}{M_h} \, \mathcal{C}_{nm}^2 \\ 
    \mathcal{C}_{nm} & \equiv \frac{1}{\lambda_{nm} R} \int U_{nm}(r,z) \, r \,  dr dz,
\end{align}
where $M_h = \rho_h \pi R^2 L$ is the mass of the $h$BAR, and $\mathcal{C}_{nm}$ is a dimensionless, $\mathcal{O}(1)$ form factor. We see that the wavelength mismatch leads to a suppression of $( M_\lambda / M_h )^2$ relative to a naive $M_{h}$ scaling, implying that a smaller $h$BAR mass is beneficial. This can also be understood physically: when a phonon is excited, the entire mass underneath the PT ($M_h$) must oscillate, which penalizes the transition probability if only a fraction of the mass, e.g., $M_\lambda$, is coupled to. This explains why using a smaller $L$ in the Gen. I stage of development is beneficial: it would remove mass which does not coherently couple to the long-wavelength signals, and weighs down the system.

The dark photon DM and high-frequency GWs do not have exactly the same interaction Hamiltonian as the previous example. However we will still find it useful to split their effects into an ``acceleration coefficient", $a_0$ (equal to $f_0 / \rho_h$ in the previous example), and a dimensionless, $\mathcal{O}(1)$ form factor which depends on the interaction details such that,
\begin{align}
    |\langle n m | \delta H_0 | 0 \rangle |^2 & \equiv a_0^2 \, \frac{2}{\omega_{nm}} \, \frac{M_\lambda^2}{M_h} \, \mathcal{C}_{nm}^2 \nonumber \\ 
    & = a_0^2 \, \frac{2 \pi \rho_h R^2}{\omega_{nm}} \, \frac{\lambda_{nm}^2}{L} \, \mathcal{C}_{nm}^2  \, .
    \label{eq:matrix_element_parameterization}
\end{align}
For each signal we will explicitly define the corresponding $\mathcal{C}_{nm}$ and $a_0$.

The phonon excitation rate from Fermi's golden rule is written in terms of the transition matrix element in Eq.~\eqref{eq:matrix_element_parameterization} as,
\begin{align}
    \Gamma_{nm} = a_0^2 \, \frac{2}{\omega_{nm}} \, \frac{M_\lambda^2}{M_h} \, \mathcal{C}_{nm}^2 \frac{4 \, \omega \, \omega_{nm} \gamma_{nm}}{(\omega^2 - \omega_{nm}^2)^2 + \omega^2 \gamma_{nm}^2} \, ,
    \label{eq:FGR}
\end{align}
where $\omega$ is the frequency of the new physics signal, and we have broadened the usual energy conserving delta function to account for the finite width of the phonons, $\gamma_{nm} \equiv \omega_{nm} / Q_\text{p}$~\cite{Mitridate:2020kly}. After around a phonon lifetime, $\tau_{nm} = 1/\gamma_{nm}$, the probability of a phonon excitation, $p_{nm}$, saturates to $p_{nm} = \Gamma_{nm} \, \tau_{nm}$.\footnote{If $Q_\text{p} > Q_\text{DM}$, where $Q_\text{DM} \sim 10^6$ is the DM ``quality factor", the excitation rate becomes limited by the DM linewidth, i.e., $\gamma_{nm} \rightarrow \omega / Q_\text{DM}$ in Eq.~\eqref{eq:FGR}. A larger $Q_\text{p}$ is still marginally beneficial since it increases $p^\text{res}$~\cite{Dixit:2020ymh} (Eq.~\eqref{eq:p_sig_on_resonance}), but also increases $p^\text{thr}$ (Eq.~\eqref{eq:psmin}) in the statistically limited regime since the number of measurements over a given observation time decreases.} On resonance $p_{nm}$ simplifies to,
\begin{align}
    p_{nm}^\text{res} = \frac{8 \, a_0^2 \, Q_{\text{p}}^2}{\omega_{nm}^3} \, \frac{M_\lambda^2}{M_h} \, \mathcal{C}_{nm}^2 \, ,
    \label{eq:p_sig_on_resonance}
\end{align}
which must be larger than the threshold value, $p_\text{thr}$, discussed in Sec.~\ref{subsec:signal_significance} to generate a detectable signal. Note that when the number of measurements is $N = T_\text{obs}/\tau_{nm}$, the condition $p_{nm}^\text{res} > p_{\text{thr}}$ is equivalent to $\Gamma_{nm} T_\text{obs} > N_\text{thr}$, where $\Gamma_{nm} T_\text{obs}$ is the expected number of phonons excited over the measurement time, and $N_\text{thr} = p_\text{thr} N$ is the dark count phonon threshold. 

A parametric understanding of how large $a_0$ must be in order to be detectable can be found by requiring that $p_{nm}^\text{res} > p_\text{thr}$, which defines an $a_0^\text{thr}$. We consider the scaling in the two different regimes, statistically limited (``stat.") and systematically limited (``sys."), identified in Eq.~\eqref{eq:p_s_diff_cases},
\begin{widetext}
    \begin{align}
        a_0^\text{thr} \sim \frac{\mu \text{m}}{\text{s}^2} \left( \frac{1}{\mathcal{C}_{nm}} \right) \left( \frac{100 \, \mu \text{m}}{R} \right) \left( \frac{L}{400 \, \mu \text{m}} \right)^{1/2} \begin{cases}
            \displaystyle  \left( \frac{\text{yr}}{T_\text{obs}} \right)^{1/4} \left( \frac{10^6}{Q_\text{p}} \right)^{3/4} \left( \frac{p_\text{dc}}{10^{-4}} \right)^{1/4} \left( \frac{\omega_{nm} / 2 \pi}{4 \, \text{GHz}} \right)^{9/4} & \text{stat.} \\[2ex] 
            \displaystyle \left( \frac{10^6}{Q_\text{p}} \right) \left( \frac{p_\text{dc}}{10^{-4}} \right)^{1/2} \left(\frac{\delta p_\text{dc} / p_\text{dc}}{10^{-4}} \right)^{1/2} \left( \frac{\omega_{nm} / 2 \pi}{4 \, \text{GHz}} \right)^{5/2}& \text{sys.} \, ,
        \end{cases} 
        \label{eq:a0_thr}
    \end{align}
\end{widetext}
where $T_\text{obs}$ is the observation time.

\begin{figure}[t!]
    \centering
    \includegraphics[width=\linewidth]{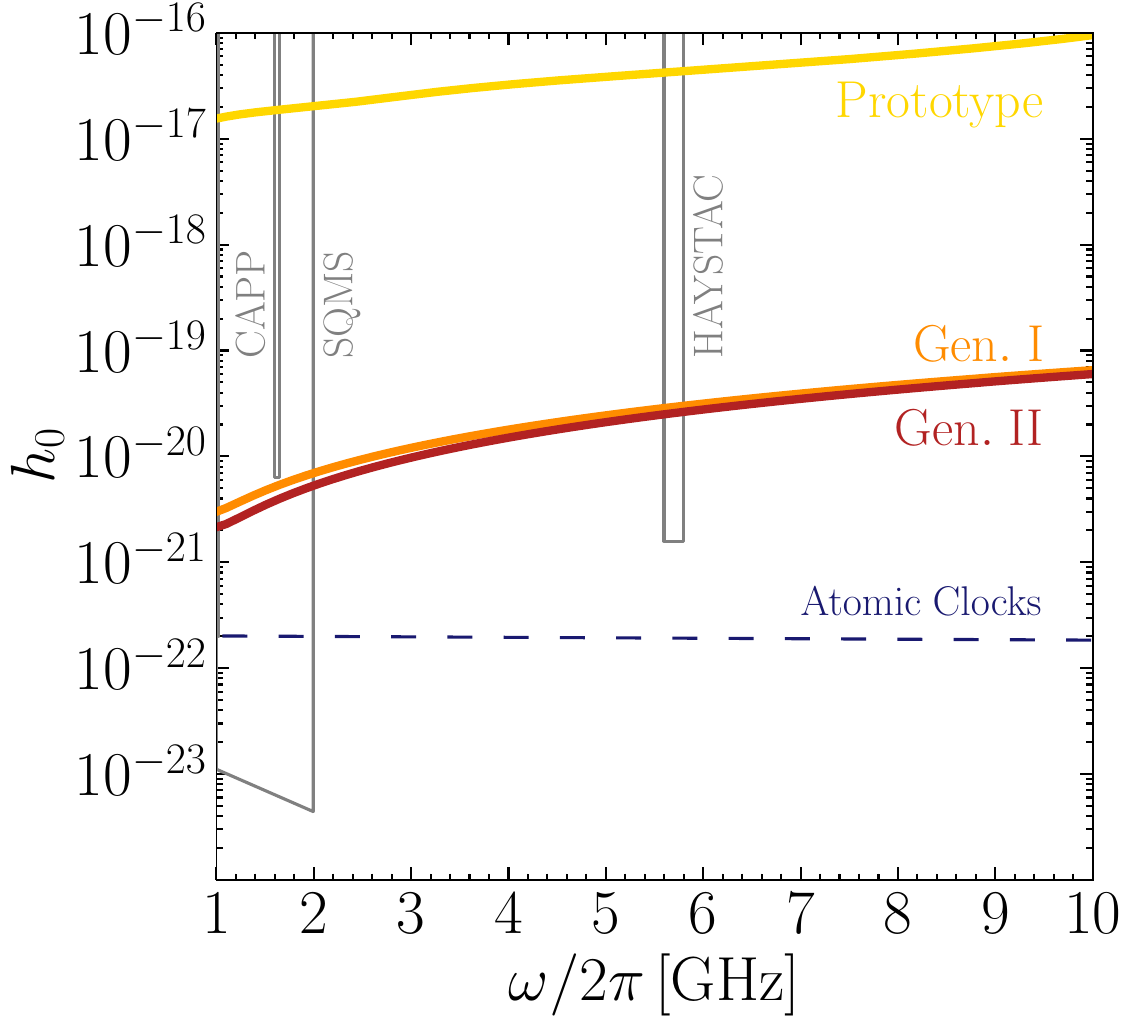}
    \caption{Projected bounds on the monochromatic, coherent GW with amplitude $h_0$ assuming $T_\text{obs} = 1\, \text{yr}$. The yellow, orange, and red lines labeled ``Prototype", ``Gen. I", and ``Gen. II", respectively, are computed assuming the parameters discussed in Table~\ref{tab:experimental_parameters}. Gray regions from current axion experiments~\cite{Berlin:2021txa} (CAPP~\cite{Lee:2020cfj}, HAYSTAC~\cite{HAYSTAC:2018rwy}, SQMS~\cite{Berlin:2021txa}) recast for $T_\text{obs} = 1 \, \text{yr}$~\cite{Kahn:2023mrj}. Also shown as a dashed line is the projection from a proposed experiment using optical atomic clocks~\cite{Bringmann:2023gba,Kahn:2023mrj} (blue, labeled ``Atomic Clocks").}
    \label{fig:h0_bounds}
\end{figure}

We now discuss two new physics signals in detail. Sec.~\ref{subsec:dp_dm} focuses on dark photon DM, and Sec.~\ref{subsec:hfgw} focuses on high-frequency GWs. Figs.~\ref{fig:dp_bounds} and~\ref{fig:h0_bounds} show the projected sensitivity on the relevant model parameters for each stage of detector development.

\subsection{Dark Photon Dark Matter}
\label{subsec:dp_dm}

To resonantly excite an $h$BAR phonon, the incoming DM mass must match the phonon frequency. For $1 \, \text{GHz} \lesssim \omega_{nm} / 2 \pi \lesssim 10 \, \text{GHz}$, this translates to DM masses, $4 \, \mu\text{eV} \lesssim m_\text{DM} \lesssim 40 \, \mu\text{eV}$. These relatively light, bosonic, DM candidates are often referred to as ``ultralight" DM. The dark photon, $V_\mu$, is a specific ultralight DM candidate~\cite{Fabbrichesi:2020wbt} which can interact with the Standard Model (SM) via a kinetic mixing,
\begin{align}
    \delta \mathcal{L} = \frac{m_V^2}{2}V_\mu V^\mu -\frac{1}{4}F'_{\mu \nu}F'^{\mu \nu} -\frac{\kappa}{2} F_{\mu \nu}' F^{\mu \nu} \, ,
    \label{eq:kinetic_mixing}
\end{align}
where $F'_{\mu \nu} = \partial_\mu V_\nu - \partial_\nu V_\mu$ is the dark photon field strength tensor, $m_V$ is the mass of the dark photon, $\kappa$ is the coupling parameter which determines the mixing strength, and $F_{\mu \nu} = \partial_\mu A_\nu - \partial_\nu A_\mu$ is the electromagnetic field strength tensor, with $A_\mu$ being the photon field. A rotation to the mass basis $A_\mu \rightarrow A_\mu - \kappa V_\mu$ generates an interaction $\kappa Q \, V_\mu \bar{\Psi} \gamma^\mu \Psi$ between $V_\mu$ and all SM fields, $\Psi$, with $U(1)_\text{EM}$ charge $Q$.

Dark photons in the $4 \, \mu\text{eV} \lesssim m_V \lesssim 40 \, \mu\text{eV}$ mass range can be produced in the early universe by a variety of different production mechanisms. In the absence of extra degrees of freedom the dark photon can be produced via quantum mechanical fluctuations during inflation~\cite{Graham:2015rva} or by the misalignment mechanism~\cite{Arias:2012az} (however the latter can become mired in technical difficulties due to the requirement of non-minimal gravitational couplings~\cite{Arias:2012az,Graham:2015rva}). Other production mechanisms are possible if the dark photon is allowed to have non-gravitational interactions with the inflaton itself~\cite{Bastero-Gil:2018uel}, or additional particles~\cite{Dror:2018pdh,Agrawal:2018vin,Co:2018lka}. Recently, in Ref.~\cite{Cyncynates:2023zwj}, it was shown that the dark photon field value at production in all of these scenarios is large enough to generate defects when the dark photon mass arises from a Higgs mechanism with perturbative quartic coupling. This places a stringent requirement on the $\kappa$, outside the reach of nearly all proposed experiments near $m_V \sim \mu\text{eV}$. However this constraint can be alleviated with further model-building; see Ref.~\cite{Cyncynates:2023zwj} for a concrete example.

Returning to the dark photon phenomenology, due to the large occupation number of dark photons within a wavelength, the dark photon field may be treated as a classical field. Furthermore, since it couples to charged particles in complete analogy with the photon, the dark photon field acts as a ``dark", spatially coherent, electric field oscillating at frequency equal to the dark photon mass, $m_V$. The amplitude of this dark electric field within the $h$BAR is $\vec{E}' \approx \kappa \sqrt{2 \rho_V} \vec{\epsilon}_V \cos{(m_V t)}/ \varepsilon_0$, where $\rho_V \approx 0.4 \, \text{GeV} / \text{cm}^3$ is the local DM mass density, $\varepsilon_0$ is the static dielectric constant screening the interaction ($\varepsilon_0 \approx 9$ for AlN~\cite{Petousis_2017}), and $\vec{\epsilon}_V$ is the dark photon polarization vector. This dark electric field will couple to the PT just as the qubit does, and therefore the interaction Hamiltonian can be written,
\begin{align}
    \delta H & = - \int d^3 \vec{x} \, e^{ijk}_\text{pt}(\vec{x}) \, \vec{E}'^i(\vec{x}, t) \, \nabla^j \vec{u}^k \nonumber \\ 
    & \approx - 2 \pi \, e_\text{pt} \, |\vec{E}'(t)| \cos{\theta_V} \nonumber \\
    & \hspace{3em} \times \int_0^R  \left[ u(r, L + h) - u(r, L) \right] \, r \,  dr \, ,
    \label{eq:H_int_dp}
\end{align}
where $\cos{\theta_V} \equiv \hat{\vec{z}} \cdot \vec{\epsilon}_V$ and we have made similar approximations as in Sec.~\ref{subsec:phonon_qubit_swap_efficiency} to simplify to the second line in Eq.~\eqref{eq:H_int_dp}. If the chip itself is piezoelectric, it will also couple to the dark electric field. However, the piezoelectric coupling in Eq.~\eqref{eq:H_int_dp} only depends on the difference of $u$ at the boundaries in the $z$ direction, so even though the chip is much larger than the PT, the dark photon-chip coupling will not be parametrically larger than the dark photon-PT coupling.

Given the interaction Hamiltonian in Eq.~\eqref{eq:H_int_dp} we find a useful definition for the phonon mode-dependent form factor, $\mathcal{C}_{nm}$, to be
\begin{align}
    \mathcal{C}_{nm} & \equiv \frac{1}{R} \int \left[ U_{nm}(r, L + h) - U_{nm}(r, L) \right] \, r \, dr \nonumber \\ 
    & = \mathcal{A}_{nm} \left[ (-1)^n - \cos{\left( \frac{n \pi L}{L + h} \right)} \right] \frac{J_1(\alpha_{nm})}{\alpha_{nm}} \, , 
    \label{eq:C_nm_dp}
\end{align}
which has the same magnitude as $\mathcal{B}_{nm}$ in Eq.~\eqref{eq:B_nm}. Using this definition and Eq.~\eqref{eq:matrix_element_parameterization}, while noting again that $\delta H_0$ is defined as $\delta H(t) \equiv \delta H_0 e^{- i \omega t} + \text{h.c.}$, the acceleration coefficient, $a_0$, for the dark photon interaction is 
\begin{align}
    a_0 & = \frac{\kappa}{\sqrt{2}} \, \frac{e_\text{pt}}{\varepsilon_0} \frac{\sqrt{\rho_V}}{\rho_h} \frac{\cos{\theta_V}}{\lambda_{nm}} \nonumber \\ 
    & \sim \frac{\mu\text{m}}{\text{s}^2} \, \cos{\theta_V} \left( \frac{\kappa}{10^{-10}} \right) \left( \frac{\omega / 2 \pi}{4 \, \text{GHz}} \right) \left( \frac{10}{\varepsilon_0} \right) \label{eq:a_0_parametric}
\end{align}
where $\lambda_{nm} = 2 \pi c_\text{l} / \omega$ is the phonon wavelength, $\rho_h \approx 3.87 \, \text{g}/\text{cm}^3$ (the density of \ce{Al2O3}), and $e_\text{pt}, c_\text{l}$ from Table~\ref{tab:experimental_parameters} were used in the parametric estimate.

\begin{figure}[t!]
    \centering
    \includegraphics[width=\linewidth]{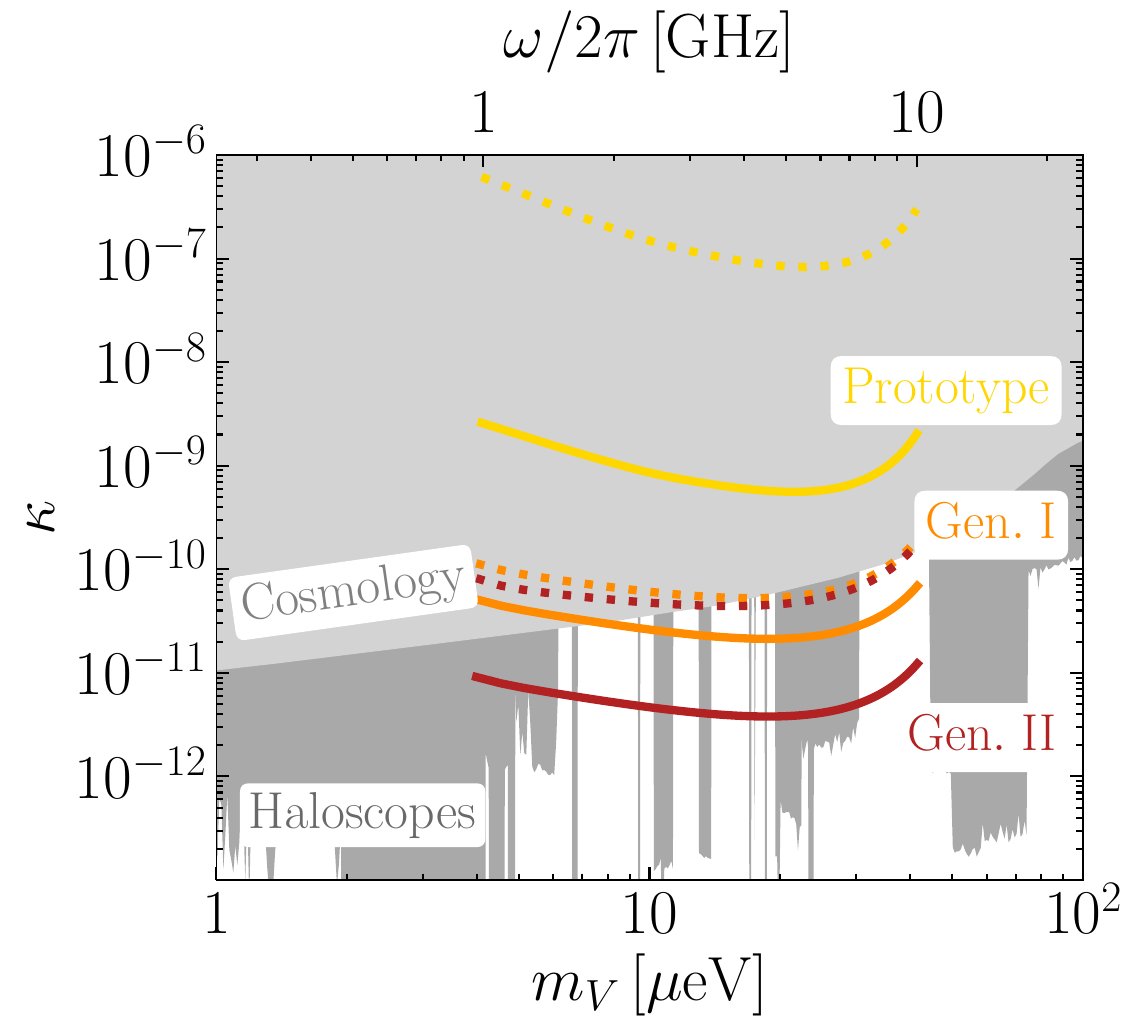}
    \caption{Projected bounds on the kinetic mixing parameter, $\kappa$ (Eq.~\eqref{eq:kinetic_mixing}), assuming $T_\text{obs} = 1 \, \text{yr}$ at a given frequency. The yellow, orange, and red lines labeled ``Prototype", ``Gen. I", and ``Gen. II", respectively, are computed assuming the parameters discussed in Table~\ref{tab:experimental_parameters}. Solid lines assume the dark photon polarization is fixed (expected in some dark photon DM production scenarios~\cite{Caputo:2021eaa} as discussed in detail in Sec.~\ref{subsec:dp_dm}), such that daily modulation can be used to remove systematics. Dashed lines assume the polarization is random which leads to no daily modulation. The light gray shaded region is excluded by cosmological probes~\cite{Arias:2012az}, and the dark gray shaded region is excluded by haloscope experiments~\cite{AxionLimits}.}
    \label{fig:dp_bounds}
\end{figure}

To compute the qc-$h$BAR sensitivity to $\kappa$, a $\cos{\theta_V}$ must be specified. If the DM polarization is distributed uniformly over polarizations then the appropriate value is $\cos{\theta_V} \rightarrow 1 / \sqrt{3}$ corresponding to an average of the rate over the three possible polarizations. However some degree of preferred polarization direction is expected in all the production scenarios discussed previously~\cite{Caputo:2021eaa}. For example, production by quantum mechanical fluctuations during inflation primarily generates the longitudinal polarization~\cite{Graham:2015rva}, production from decay of additional particles generates a single transverse polarization~\cite{Bastero-Gil:2018uel}, production via the misalignment mechanism fixes a specific field direction~\cite{Arias:2012az}, and the mechanism in Ref.~\cite{Cyncynates:2023zwj} generates primarily transverse polarizations. Whether this directionality persists for cosmologically long times is currently unknown, and beyond the scope of this work, but Ref.~\cite{Caputo:2021eaa} estimates that the directionality can survive for long times in weak gravitational potentials. Assuming the dark $E$ field polarization is fixed in the Galactic frame leads to a time-dependent $\cos{\theta_V(t)}$ in the Earth frame, which oscillates over a day. The directional dependence of the detector can then be used to remove systematics, since backgrounds are not expected to fluctuate on the specific time scale of a day. This is a scenario in which a non-differential measurement scheme, discussed in Sec.~\ref{subsec:signal_significance}, can be used.\footnote{To implement this search strategy the counts data must first be binned on timescales small enough to see the modulation, e.g., in $\sim 1 \, \text{hr}$ bins. This binned data can then be substituted in to a test-statistic whose distribution defines a meaning of statistical significance, as done previously in Refs.~\cite{Griffin:2018bjn,Coskuner:2021qxo}. Following a similar procedure we find that even if $p_\text{dc}$ fluctuates at the $10 \%$ level between measurements, the signal rate still only needs to be larger than the statistical fluctuations.}

In Fig.~\ref{fig:dp_bounds} we show the projected bounds on $\kappa$ for the three stages of detector development (labeled ``Prototype", ``Gen. I" and ``Gen. II", assuming the relevant parameters in Table~\ref{tab:experimental_parameters}) and $T_\text{obs} = 1\, \text{yr}$. Solid lines assume that modulation can be used to remove systematics (the threshold signal probability is given by Eq.~\eqref{eq:non_diff}) and $\cos{\theta_V}$ varies maximally over the day. Dotted lines assume there is no modulation (the threshold signal probability is given by Eq.~\eqref{eq:psmin}), and $\cos{\theta_V} \rightarrow 1 / \sqrt{3}$. Additionally we show bounds from cosmological observables (light gray, ``Cosmology"), specifically due to CMB distortions in this mass range~\cite{Arias:2012az}, and those from other haloscope-type experiments (dark gray, ``Haloscopes", including Dark E-field~\cite{Levine:2024noa}, WISPDMX~\cite{Nguyen:2019xuh}, ADMX~\cite{ADMX:2010ubl}, FAST~\cite{An:2022hhb}, CAPP~\cite{Lee:2020cfj}, SQMS~\cite{Cervantes:2022gtv}, HAYSTAC~\cite{HAYSTAC:2020kwv}, ORGAN~\cite{McAllister:2017lkb}, SQuAD~\cite{Dixit:2020ymh}, APEX~\cite{APEX:2024jxw}, QUALIPHIDE~\cite{Ramanathan:2022egk}, SUPAX~\cite{Schneemann:2023bqc}, QUAX~\cite{Alesini:2020vny}, GigaBREAD~\cite{BREAD:2023xhc}, BRASS~\cite{Bajjali:2023uis}, ORPHEUS~\cite{Cervantes:2022yzp}, and MADMAX~\cite{MADMAX:2024jnp}), compiled with the help of Ref.~\cite{AxionLimits}. The non-trivial $m_V$, or $\omega$, dependence is due to the $\mathcal{C}_{nm}$ form factor in Eq.~\eqref{eq:C_nm_dp}. Since this form factor also appears in the calculation of $g_{nm}$ (Eq.~\eqref{eq:couplings}), the shape of the bounds on $\kappa$ in Fig.~\ref{fig:dp_bounds} can be understood from the left panel of Fig.~\ref{fig:swap_efficiency}.

Other vector DM candidates, such as those arising in theories with spontaneously broken $U(1)_B$ or $U(1)_{B - L}$ symmetries, can also couple to phonons in the $h$BAR as $\delta H \sim g \int d^3 \vec{x} \, q(\vec{x}) \, \vec{E}'(\vec{x}) \cdot \vec{u}(\vec{x})$, where $g, q(\vec{x})$ are the relevant gauge coupling and charge density, respectively. If the PT and bulk target chip have the same $q$, then for long-wavelength $\vec{E}'$ the interaction Hamiltonian vanishes for the phonon modes considered here. If this selection rule was not present (e.g., if the PT and bulk target chip had dramatically different $B$ or $B - L$ charges, or if the boundary conditions in the $\hat{\vec{z}}$ direction were different), the acceleration coefficient would parameterically be $a_0 \sim g (q / \rho_h) \sqrt{\rho_V / 2} \sim \mu \text{m} / \text{s}^2 \, (g / 10^{-18})$. However even in this scenario the existence of a vector particle coupling via $U(1)_B$ or $U(1)_{B - L}$ with $g \sim 10^{-18}$ is already ruled out down to $g \sim 10^{-22}$ at $m_V \sim \mu\text{eV}$ and $g \sim 10^{-20}$ at $m_V \sim 100 \, \mu\text{eV}$ by short range tests of the equivalence principle~\cite{Smith:1999cr,Arvanitaki:2015iga,Hees:2018fpg} and fifth force experiments~\cite{Adelberger:2009zz,Arvanitaki:2015iga,Hees:2018fpg}. Similar conclusions apply to the scalar DM models considered in Ref.~\cite{Arvanitaki:2015iga}, which have the additional disadvantage of no modulation effects at leading order.

\subsection{High-Frequency Gravitational Waves}
\label{subsec:hfgw}

Phonons within acoustic resonators have long been used to search for GWs. The original Weber bar~\cite{Weber:1960zz} and modern resonant mass detectors~\cite{Forward:1971mel,Hamilton:1989nb,Johnson:1993cr,Harry:1996gh,Lobo:1999ns,Gottardi:2007zn,Aguiar:2010kn,DaSilvaCosta:2014trv} use $\omega / 2 \pi \sim \text{kHz}$ frequency acoustic modes. Bulk acoustic resonator devices operate at even higher frequencies, typically in the $\text{MHz} \lesssim \omega / 2 \pi \lesssim \text{GHz}$ frequency range~\cite{Goryachev:2014yra,Goryachev:2021zzn,Tobar:2023ksi,Campbell_2023}. Recently it has also been shown that direct detection experiments sensitive to single optical phonon excitations are also sensitive to GWs in the $\text{THz} \lesssim \omega / 2 \pi \lesssim 100 \, \text{THz}$ frequency range~\cite{Kahn:2023mrj}. The $h$BAR device discussed here thus complements these other searches, operating in the frequency range above traditional bulk acoustic resonators, but below direct detection experiments. GWs above the $\text{MHz}$ frequency are commonly referred to as ``high-frequency" GWs; see Refs.~\cite{Aggarwal:2020olq,Domcke:2023qle} for recent reviews on other detection methods for high-frequency GWs.

High-frequency GWs may be generated by a variety of different cosmological and astrophysical sources~\cite{Aggarwal:2020olq}. Cosmological sources, or those from the early universe, generate a stochastic gravitational wave background, whose presence at $\text{nHz}$ frequencies has recently been evidenced by pulsar timing arrays, e.g., NANOGrav~\cite{NANOGrav:2023gor}. GWs generated in the early universe are severely constrained by the CMB and BBN, which limit the energy density in relativistic degrees of freedom to $\Omega_\text{GW} \lesssim 10^{-6}$~\cite{Cyburt:2015mya,Aggarwal:2020olq}. This limits the characteristic strain to $h \sim 10^{-30} \left( \text{GHz} / \nu \right) \left( \Omega_\text{GW} / 10^{-6} \right)^{1/2}$~\cite{Kahn:2023mrj}, well outside of the projected reach of any future $h$BAR device. However, high-frequency GWs can also be generated in the late universe via superradiance~\cite{Arvanitaki:2010sy,Arvanitaki:2014wva,Brito:2015oca,Baryakhtar:2020gao}, or the inspiral of light compact objects, e.g., primordial black holes~\cite{Maggiore:2007ulw}. These sources can have amplitudes that are substantially larger. For example, in a superradiant annihilation process, the GW amplitude may be modelled as $h_0(t) \approx h_s / (1 + t / \tau_\text{sr})$~\cite{Kahn:2023mrj}, where $\tau_\text{sr}$ is the superradiance coherence time, and $h_s \sim 10^{-23} \left( 10 \, \text{kAU} / r \right) \left( 10 \, \mu \text{eV} / \omega \right)$ for a compact object with mass $M_\text{CO} \sim M_\text{pl}^2 / \omega \sim 10^{-5} M_\odot \left( 10 \, \mu \text{eV} / \omega \right)$. While $h_s$ cannot be directly compared to the $h_0$ shown in Fig.~\ref{fig:h0_bounds}, since $\tau_\text{sr} \sim 30 \, \text{s} \, \left(10 \, \mu \text{eV} / \omega \right)$ is much smaller than a year, such a signal has a much larger amplitude than that of a stochastic, cosmological source, if it is close enough to the Earth.

The effect of an incoming GW with frequency $\omega$, whose wavelength is much longer than the size of the detector, is to induce a strain which has an interaction Hamiltonian given by~\cite{Maggiore:2007ulw,Tobar:2023ksi,Kahn:2023mrj},
\begin{align}
    \delta H & = \frac{1}{2} \int d^3 \vec{x} \, \rho_h \, \ddot{h}^{ij}(\vec{x}, t) \, \vec{x}^i \, \vec{u}^j \nonumber \\ 
    & \approx -\frac{\omega^2 \rho_h \, e^{zz}_\text{GW} }{2} \, h_0 \cos{(\omega t)}  \int z \, u \, d^3\vec{x}
    \label{eq:H_int_hfgw}
\end{align}
where $\ddot{h}^{ij} = -\omega^2 h_0 \, \vec{e}^{ij}_\text{GW} \cos{(\omega t)}$ is the second time derivative of the dimensionless GW amplitude in the TT frame, $e^{zz}_\text{GW} \equiv \vec{e}^{ij}_\text{GW} \hat{\vec{z}}^i \hat{\vec{z}}^j$ is the GW polarization tensor, $\vec{e}^{ij}_\text{GW}$, projected onto the $\hat{\vec{z}}$ direction, and in the second line we have made the appropriate simplifications for the $h$BAR device discussed in Sec.~\ref{sec:DeviceDesignAndOperation}. We define the phonon mode-dependent form factor, $\mathcal{C}_{nm}$, as
\begin{align}
    \mathcal{C}_{nm} & \equiv \frac{\pi^2}{\lambda_{mn}^2 R} \int z \, U_{nm}(r, z) \, r \, dr \, dz \\ 
    & \hspace{-1.5em} = \mathcal{A}_{nm} \left[ (-1)^n - 1 \right]  \left[ \frac{J_1(\alpha_{nm})}{\alpha_{nm}} + \frac{J_0(\alpha_{nm})}{K_0(\beta_{nm})} \frac{K_1(\beta_{nm})}{\beta_{nm}} \right] \, ,
    \label{eq:Cnm_h0}
\end{align}
where we note that $n$ must be odd for nonzero coupling to the GW. As in Sec.~\ref{subsec:dp_dm}, the corresponding acceleration coefficient, $a_0$, is found by combining Eq.~\eqref{eq:Cnm_h0} and Eq.~\eqref{eq:matrix_element_parameterization}, noting again that $\delta H_0$ is defined as $\delta H(t) \equiv \delta H_0 e^{- i \omega t} + \text{h.c.}$,
\begin{align}
    a_0 = \frac{e_\text{GW}^{zz} \, c_\text{l} \, \omega}{2 \pi} \, h_0 \sim 0.4 \, \frac{\mu \text{m}}{\text{s}^2} \, e^{zz}_\text{GW} \, \left( \frac{\omega / 2 \pi}{4 \, \text{GHz}} \right) \left( \frac{h_0}{10^{-20}} \right) \, .
    \label{eq:gw_a_0_parametric}
\end{align}

As in Sec.~\ref{subsec:dp_dm}, in Fig.~\ref{fig:h0_bounds} we show the projected bounds for the three stages of detector development (labeled ``Prototype", ``Gen. I" and ``Gen. II", assuming the relevant parameters in Table~\ref{tab:experimental_parameters}) and $T_\text{obs} = 1 \, \text{yr}$, assuming a monochromatic GW source with frequency $\omega$ and amplitude $h_0$. While this idealized GW source makes comparing experiments simple, the candidate sources of GWs at these frequencies, discussed previously, will typically not emit coherently over year-long timescales, meaning that modulation of the signal cannot be utilized. Therefore the projected bounds on $h_0$ assume the differential measurement scheme is used, and includes systematic uncertainties. Furthermore we take $e^{zz}_\text{GW} \rightarrow \sqrt{2 / 15}$ which angularly averages over incoming polarizations and directions.\footnote{The angularly averaged quantity is $(e^{zz}_\text{GW})^2=\hat{\vec{z}}^i \hat{\vec{z}}^j \hat{\vec{z}}^k \hat{\vec{z}}^l \int d \Omega \,  \left[ \frac{1}{2} \sum_{\lambda} e_{\lambda}^{ij} e_{\lambda}^{kl} \right] / 4 \pi$ and the polarization sum is simplified using the completeness relation, $\frac{1}{2} \sum_\lambda e_{\lambda}^{ij} e_{\lambda}^{kl} = \left( P^{ik} P^{jl} + P^{il} P^{jk} - P^{ij} P^{kl}\right)/4$, and $P^{ij} = \delta^{ij} - \hat{\vec{q}}^i \hat{\vec{q}}^j$~\cite{Ghiglieri:2020mhm}.} We also show bounds utilizing the inverse Gertsenshtein effect in haloscope experiments (CAPP~\cite{Lee:2020cfj}, HAYSTAC~\cite{HAYSTAC:2018rwy}, SQMS~\cite{Berlin:2021txa}), as well as projected bounds as dashed lines from atomic clocks~\cite{Bringmann:2023gba} (blue). Dielectric haloscope experiments, e.g., MADMAX~\cite{Majorovits:2016yvk,Millar:2016cjp,MADMAX:2019pub}, have also been recently shown to have sensitivity to high-frequency GWs in this frequency range~\cite{Domcke:2024eti}.

\section{Discussion}
\label{sec:Conclusions}

To detect weakly coupled new physics, such as ultralight dark photon DM or high-frequency GWs, extremely sensitive detectors are required. At the smallest interaction strengths, these phenomena may only produce single quanta within a detector, and therefore leveraging recent advancements in quantum sensing is crucial. In this paper, we explore how a high-overtone bulk acoustic resonator ($h$BAR) coupled to a superconducting qubit can precisely detect individual phonon quanta, complementing haloscope based approaches also searching for these elusive signals.

In Sec.~\ref{sec:DeviceDesignAndOperation} we detailed the detector design and operation, illustrated schematically in Fig.~\ref{fig:DeviceDesignDiagram}, and discussed the phonon modes in the $h$BAR shown in Fig.~\ref{fig:phonon_mode_illustration} (derived in App.~\ref{app:hbar_eigenmodes}). In Sec.~\ref{sec:DevicePerformance} we detailed the detector performance, first computing the optimal phonon-qubit swap efficiency in Sec.~\ref{subsec:phonon_qubit_swap_efficiency} (shown in Fig.~\ref{fig:swap_efficiency}), and then discussed the relevant backgrounds in Sec.~\ref{subsec:Backgrounds} (summarized in Fig.~\ref{fig:frequency_dependence_of_pdc}). Lastly in Sec.~\ref{sec:DarkMatterSearchModes} we discussed how dark photon DM can couple to the $h$BAR phonons pizeoelectrically (Sec.~\ref{subsec:dp_dm}) and how high-frequency GWs can excite phonons due to the induced strain (Sec.~\ref{subsec:hfgw}), and then derived sensitivity to both of these new physics signals for each stage of detector development (Figs.~\ref{fig:dp_bounds} and~\ref{fig:h0_bounds}, respectively). The ``Gen. II" stage detector, which includes a more optimized detector geometry and optimistic background estimates, could complement the variety of haloscopes also searching for these signals. Such a detector represents a long-term goal for this technology, and we have quantified the parameters needed to be optimized to achieve this goal.

While we have focused on the sensitivity of the qc-$h$BAR to resonant new physics signals, the device architecture may also be sensitive to light, sub-GeV DM scattering. Single phonon excitations in bulk crystal targets have been shown to be exceptionally sensitive to DM scattering~\cite{Knapen:2017ekk,Griffin:2018bjn,Trickle:2019nya,Kurinsky:2019pgb,Cox:2019cod,Coskuner:2021qxo,Trickle:2020oki,Griffin:2019mvc,Knapen:2021bwg,Griffin:2020lgd,Taufertshofer:2023rgq,Raya-Moreno:2023fiw}. Previous literature has focused on the excitation of single phonons in the $1 \, \text{meV} \lesssim \omega \lesssim 100 \, \text{meV}$ energy range since typical transition edge sensor detection technology has $\mathcal{O}(\text{meV})$ thresholds (see, for example, the TESSERACT experimental proposal~\cite{Chang2020}). Since the threshold of the $h$BAR device discussed here is much lower, $\mathcal{O}(10 \, \mu\text{eV})$, it is sensitive to not only DM directly exciting $\mathcal{O}(10 \, \mu\text{eV})$ energy phonons, but also higher energy phonons which decay, or downconvert, to the $\mathcal{O}(10 \, \mu\text{eV})$ energy range. Utilizing low-energy phonons as secondaries is also a strategy employed by other direct detection experiments, e.g., SuperCDMS~\cite{SuperCDMS:2024yiv}, where the initial energy injection from the DM produces relatively high energy electron excitations, which then create phonons via the Neganov-Trofimov-Luke effect~\cite{Neganov:1985khw,Luke:1988xcw}. To understand the sensitivity of a qc-$h$BAR to these signals one needs to know the energy spectrum of the resulting phonons as a function of time. While we plan to explore this energy spectrum in detail in future work, we briefly comment on some promising paths forward.

Generally, bulk anharmonic downconversion is too slow to downconvert athermal phonons to $\mathcal{O}(\mu \text{eV})$ energies before they escape the chip through mounting structures~\cite{HarrelsonPhononDecoherence,Hernandez:2024tfn,McEwen}. However, the downconversion rate can be enhanced by a variety of processes, e.g., isotopic scattering through decreasing material purity, engineering of substrate surfaces for surface-based downconversion, or addition of normal metal or superconducting metal films to surfaces for in-film downconversion~\cite{Downconversion_2022}.

Of these strategies we consider superconducting film downconversion to be particularly promising. Athermal phonons incident on a low-\(T_{c}\) superconducting thin film will break Cooper pairs into quasiparticles, which then quickly relax to the superconducting gap $\Delta$~\cite{Kaplan,McEwen,FischerPhotonMediatedRecombination}. This relaxation process is generally faster than $\mathcal{O}(\mu$s) if $\Delta$ is much smaller than the initial phonon energy. As they near the gap, these quasiparticles recombine to produce phonons with energy 2$\Delta$. By placing this superconducting film near the $h$BAR, and selecting a material with $2 \Delta$ close to the $h$BAR phonon energies, a ``resonant" conversion of meV-scale to $h$BAR phonons could be achieved. Low-gap materials that may enable this include Hf (\(2\Delta\approx30 \, \mu\)eV)~\cite{HafniumCriticalTemp} and \(\alpha\)-W (\(2\Delta\approx3 \, \mu\)eV)~\cite{TungstenCriticalTemp} or a tuned combination of \(\alpha\)-W and \(\beta\)-W~\cite{JastramTunableTc}, which allows access to a broad range of superconducting gaps. The usefulness of this strategy depends on how well the phonons emitted from the superconducting thin film convert to $h$BAR phonons. If this resonant conversion is possible without negatively impacting qc-$h$BAR performance, it would provide a means of accessing a much broader range of new physics signals and lowering detector energy threshold down toward the $10~\mu$eV scale.

The qc-$h$BAR is also flexible to applied readout scheme. For example, a potential approach to improving the readout scheme might be operating the coupled qubit-resonator system in the strong dispersive regime which enables the possibility of quantum non-demolition (QND) readout of the resonator state. These methods have found significant use in electromagnetic resonators, especially in the context of detection of axions and dark photons \cite{schuster2007resolving,Dixit:2020ymh}. Recently there have also been demonstrations of operating acoustic resonators, including one consisting of an \hbarName{} type device, in the strong dispersive regime where the detuning between the qubit and the resonator frequency far exceeds the coupling strength between them \cite{von2022parity, lee2023strong}. This allows for the QND measurement of the parity of the phonon number which induces a frequency shift of the qubit, through an interferometric Ramsey readout, and might be beneficial for detection of events generating a single phonon. 

\begin{acknowledgments}
    We would like to thank Kim Berghaus, Asher Berlin, Edgard Bonilla, Enectali Figueroa-Feliciano, Roni Harnik, Stefan Knirck, Paolo Privitera, Sara Sussman, Dylan Temples, Yikun Wang, and Kathryn Zurek for helpful discussion. We would also like to thank Daniel Baxter and Rakshya Khatiwada for thoughtful conversations and feedback on the draft. This manuscript has been authored by Fermi Research Alliance, LLC under Contract No. DE-AC02-07CH11359 with the U.S. Department of Energy, Office of Science, Office of High Energy Physics. This work was supported by the U.S. Department of Energy, Office of Science, National Quantum Information Science Research Centers, Quantum Science Center and the U.S. Department of Energy, Office of Science, High Energy Physics Program Office. TL was supported by Department of Energy grant DE-SC0022104. MS's work was supported by the US Department of Energy Office of Science under Award No. DE-SC0022104 and No. DE-SC0009919, the Research Network Quantum Aspects of Spacetime (TURIS), and funded/co-funded by the European Union (ERC, NLO-DM, 101044443). CRC and ANC are supported by the Defense Advanced Research Projects Agency (DARPA) under Agreement No. HR00112490364; the Air Force Office of Scientific Research (AFOSR grant FA9550-20-1-0364);  the Army Research Office (ARO grant W911NF2310077), and in part by the U.S. Department of Energy Office of Science National Quantum Information Science Research Center Q-NEXT; UChicago's MRSEC (NSF award DMR-2011854);  and the NSF QLCI for HQAN (NSF award 2016136).
\end{acknowledgments}

\appendix
\onecolumngrid

\section{\texorpdfstring{$h$}{text}BAR Eigenmodes}
\label{app:hbar_eigenmodes}

In this appendix we derive the phonon eigenmodes of the \hbarName{} discussed in Sec.~\ref{sec:DeviceDesignAndOperation}, and discuss the approximations needed to find the analytic solutions in Eq.~\eqref{eq:radial}. We refer the reader to Refs.~\cite{10.1121/1.1912835,Goryachev:2014yra,hBAR_masters} for similar derivations in systems with different geometries and boundary conditions. We begin by assuming that the mechanical properties, e.g., sound speeds and mass densities, of the PT and bulk are identical, so that the elastic wave equation only needs to be solved in one domain. Furthermore we assume that the $h$BAR material is isotropic, and therefore the elastic wave equation is parameterized by just the transverse, $c_\text{t}$, and longitudinal, $c_\text{l}$, speeds of sound~\cite{Landau:1986aog},
\begin{align}
    \frac{ \partial^2 \vec{u} }{\partial t^2} = c_\text{t}^2 \, \nabla \times \nabla \times \vec{u} + c_\text{l}^2 \, \nabla (\nabla \cdot \vec{u}) \, .
    \label{eq:elastic_wave_equation_isotropic_app_0}
\end{align}
To focus on eigenmodes with displacement in the $\hat{\vec{z}}$ direction, $\vec{u} = u \vec{\hat{z}}$, we must assume that fluctuations in the $\hat{\vec{x}}, \hat{\vec{y}}$ directions are small, $\partial u / \partial x \sim \partial u / \partial y \ll \partial u / \partial z$, to remove mixing between the components of $\vec{u}$. Under this assumption we can recast Eq.~\eqref{eq:elastic_wave_equation_isotropic_app_0} in cylindrical coordinates $(r, \phi, z)$ as,
\begin{align}
    \frac{ \partial^2 u }{\partial t^2} & = c_\text{l}^2 \frac{\partial^2 u}{\partial z^2} + c_\text{t}^2  \left( \frac{\partial^2 u}{\partial r^2} + \frac{1}{r} \frac{\partial u}{\partial r} + \frac{1}{r^2} \frac{\partial^2 u}{\partial \phi^2} \right) \, .
    \label{eq:elastic_wave_equation_isotropic_app_1}
\end{align}
We will find it useful to work in the dimensionless variables,
\begin{align}
    \rho \equiv r / R~~~,~~~\tau \equiv c_\text{l} t / L~~~,~~~\lambda^2 \equiv \frac{c_\text{t}^2}{c_\text{l}^2} \frac{L^2}{R^2} \, ,
\end{align}
where, as in the main text, $L$ is the length of the \hbarName{} in the $\hat{\vec{z}}$ direction, and $R$ is the radius of the PT. In these dimensionless variables Eq.~\eqref{eq:elastic_wave_equation_isotropic_app_1} simplifies to,
\begin{align}
    \frac{ \partial^2 u }{\partial \tau^2} & = L^2 \frac{\partial^2 u}{\partial z^2} + \lambda^2 \left( \frac{\partial^2 u}{\partial \rho^2} + \frac{1}{\rho} \frac{\partial u}{\partial \rho} + \frac{1}{\rho^2} \frac{\partial^2 u}{\partial \phi^2} \right) \, .
    \label{eq:elastic_wave_equation_isotropic_app_2}
\end{align}

To solve Eq.~\eqref{eq:elastic_wave_equation_isotropic_app_2} we must specify the boundary conditions. We assume the \hbarName{} has stress-free boundary conditions on the top and bottom surfaces in the $\hat{\vec{z}}$ direction. Defining the bottom surface at $z = 0$, and the top surface is at $z \equiv L + h(\rho) \equiv L \, ( 1 + \Delta(\rho) )$, where $\Delta \ll 1$, the normal components of the stress tensor, $\bm{\sigma}$, must satisfy,
\begin{align}
    \hat{\vec{z}} \cdot \bm{\sigma}|_{z = 0} & = \hat{\vec{z}} \cdot \bm{\sigma}|_{z = L + h(\rho)} = 0 \label{eq:stress_free}\, .
\end{align}
Strictly speaking these cannot be exactly satisfied for solutions where $\vec{u}$ has only a $\hat{\vec{z}}$ component. However, when the $\hat{\vec{x}}$ and $\vec{\hat{y}}$ components, and their derivatives, are small compared to the $\hat{\vec{z}}$ component and its derivatives, at leading order the boundary conditions in Eqs.~\eqref{eq:stress_free} can be satisfied by requiring,
\begin{align}
    \frac{\partial u}{\partial z}\bigg|_{z = 0} & = \frac{\partial u}{\partial z}\bigg|_{z = L + h(\rho) } = 0 \, .
    \label{eq:z_boundary_cartesian}
\end{align}

Solving the elastic wave equation with position dependent boundary conditions is generally difficult. However it simplifies greatly if we change variables,
\begin{align}
    \zeta \equiv \frac{z}{L + h(\rho)} = \frac{z}{L} \frac{1}{1 + \Delta(\rho)} \, ,
    \label{eq:zeta_definition}
\end{align}
and assume that $h(\rho)$ is slowly varying in $r, \phi$ such that 
\begin{align}
    L^2 \frac{\partial^2 u}{\partial z^2} \approx \left( 1 + \Delta \right)^{-2} \frac{\partial^2 u}{\partial \zeta^2} \approx \left( 1 - 2 \Delta \right) \frac{\partial^2 u}{\partial \zeta^2} \, .
    \label{eq:diff_transformation_approx}
\end{align}
While the specific height profile used in the main text, $\Delta(\rho) = (h/L) \, \Theta(1 - \rho)$, does not strictly satisfy the slowly varying requirement, if $h/L \ll 1$ these corrections are expected to be small. Using Eq.~\eqref{eq:diff_transformation_approx} transforms the elastic wave equation (Eq.~\eqref{eq:elastic_wave_equation_isotropic_app_2}) and boundary conditions (Eq.~\eqref{eq:stress_free}) to,
\begin{align}
    \frac{ \partial^2 u }{\partial \tau^2} & = (1 - 2 \Delta) \frac{\partial^2 u}{\partial \zeta^2} + \lambda^2 \left( \frac{\partial^2 u}{\partial \rho^2} + \frac{1}{\rho} \frac{\partial u}{\partial \rho} + \frac{1}{\rho^2} \frac{\partial^2 u}{\partial \phi^2} \right) \, .\label{eq:elastic_wave_equation_isotropic_app_3}\\ 
    \frac{\partial u}{\partial \zeta}\bigg|_{\zeta = 0} & = \frac{\partial u}{\partial \zeta}\bigg|_{\zeta = 1} = 0 \label{eq:elastic_wave_equation_isotropic_boundary}\, .
\end{align}

\begin{figure}[ht!]
    \centering
    \includegraphics[width=\linewidth]{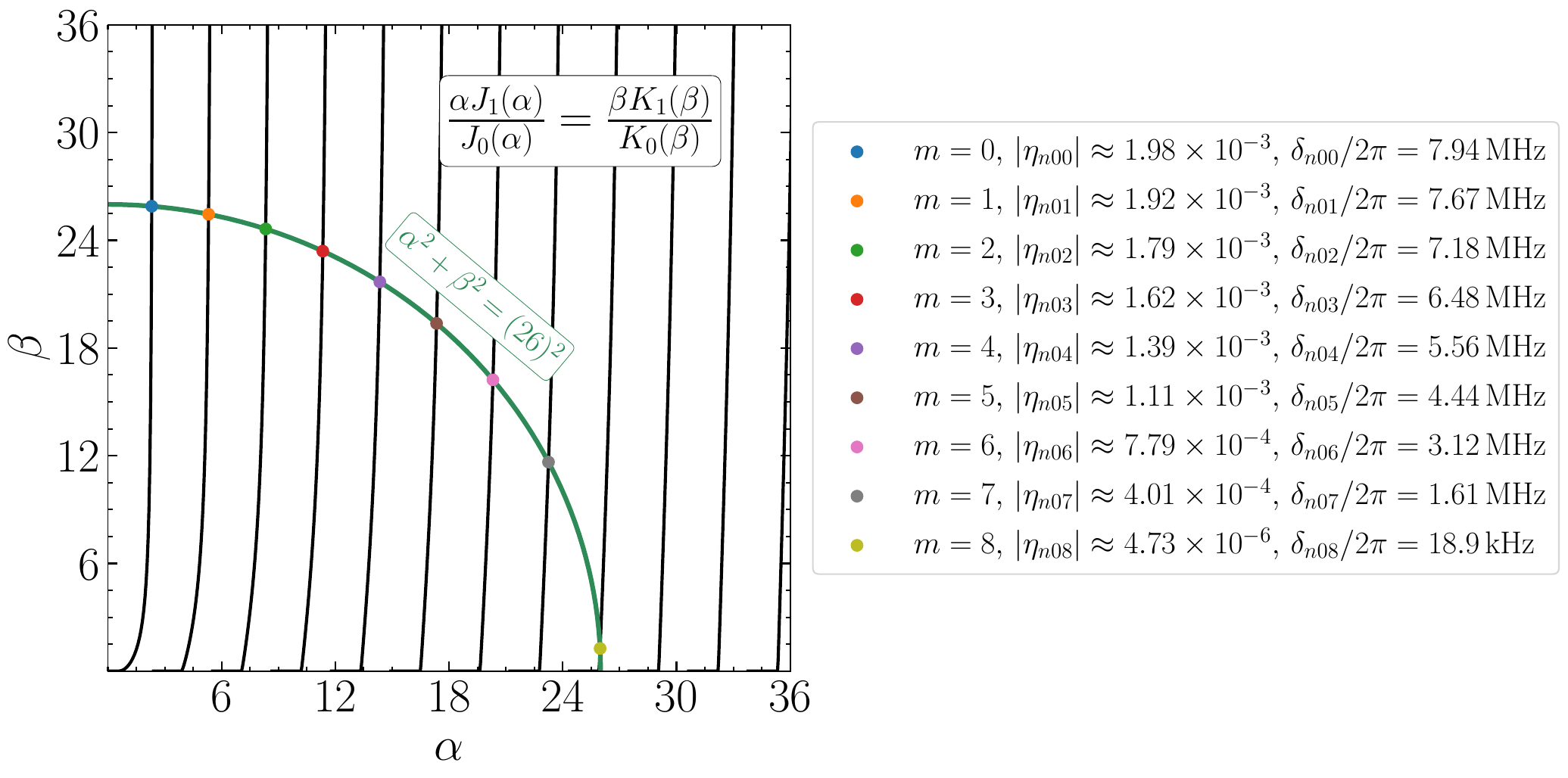}
    \caption{Graphical representation of the $\ell = 0$ solutions to Eq.~\eqref{eq:bound_solve}. The black and green lines correspond to the first and second equations in Eq.~\eqref{eq:bound_solve}, respectively, and the intersections are solutions. For the prototype development stage parameters in Table~\ref{tab:experimental_parameters}, and $\omega / 2 \pi \approx 4 \, \text{GHz}$, $\chi = (q / \lambda) \sqrt{2 \delta} \approx (\omega R / c_\text{t}) \sqrt{2 h / L} \approx 26$. Larger values of $\chi$ will lead to more bound phonon mode solutions. For large $\alpha$, the black lines go vertical at, approximately, multiples of $\pi$. As in Fig.~\ref{fig:phonon_mode_illustration}, the mode with frequency $\omega / 2 \pi \approx 4 \, \text{GHz}$ in the prototype development stage corresponds to $n = 291$. In the legend we present the $\eta_{n 0 m}$ and detuning $\delta_{n 0 m} \equiv \omega_n - \omega_{n 0 m}$, where $\omega_n = n \pi c_\text{l} / L$, corresponding to each of the solutions.}
    \label{fig:app_bound_mode_solutions}
\end{figure}

We now look for solutions of the form,
\begin{align}
    u(\rho, \phi, \zeta, \tau; q) & = \psi(\rho, \phi, \zeta; q) \, e^{\pm i q (\zeta - \tau)} \, ,
    \label{eq:parameterized_solutions}
\end{align}
which are parameterized by the positive variable $q$. Note that identical solutions can be found if $\tau \rightarrow - \tau$ in Eq.~\eqref{eq:parameterized_solutions}, which can then be combined to create real solutions in time. This decomposition is useful because it splits $u$ in to a term which varies rapidly in $z$, $e^{\pm i q \zeta}$, and a term which varies slowly in $z$, $\psi$, where slowly varying means $q \gg \partial \psi / \partial \zeta$. Substituting Eq.~\eqref{eq:parameterized_solutions} into Eq.~\eqref{eq:elastic_wave_equation_isotropic_app_3} gives,
\begin{align}
    \pm i \frac{\partial \psi}{\partial \zeta} = - \frac{\lambda^2}{2 q} \left( \frac{\partial^2 \psi}{\partial \rho^2} + \frac{1}{\rho} \frac{\partial \psi}{\partial \rho} + \frac{1}{\rho^2} \frac{\partial^2 \psi}{\partial \phi^2} \right) - q \, \Delta \, \psi \, ,
    \label{eq:eom_approx_1}
\end{align}
where we have ignored the $\partial^2 \psi/\partial \zeta^2$ term which is small because $\psi_q$ varies slowly in $z$. To gain some intuition for Eq.~\eqref{eq:eom_approx_1}, notice that the solutions with $+ i$ on the left-hand side satisfy the Schr\"odinger equation where $\zeta$ is the analog of time, $q / \lambda^2$ is an effective mass, and $- q \Delta$ is an effective potential. Notably this effective potential is directly determined by the height profile of the surface, $\Delta$. Considering the height profile from the main text, $\Delta = \delta \, \Theta(1 - \rho)$, where $\delta \equiv h / L$, Eq.~\eqref{eq:eom_approx_1} can be solved with separation of variables,
\begin{align}
    \psi_{\ell}(\rho, \phi, \zeta; q, \eta) = R_{\ell}(\rho; q, \eta) \, e^{i \ell \phi} \, e^{\mp i q \eta \zeta} \, ,
    \label{eq:decomposition}
\end{align}
where $\ell$ is an integer, and $\eta$ is another undetermined parameter. Solutions real in $\phi$ can be constructed from the $\ell \rightarrow -\ell$ solutions of Eq.~\eqref{eq:decomposition}. Substituting Eq.~\eqref{eq:decomposition} in to Eq.~\eqref{eq:eom_approx_1} generates the differential equation for $R_{\ell}$,
\begin{align}
    \rho^2 \frac{\partial^2 R_{\ell}}{\partial \rho^2} + \rho \frac{\partial R_{\ell}}{\partial \rho}  + \left( 2 \frac{q^2 \rho^2 }{ \lambda^2 } ( \eta + \Delta ) - \ell^2 \right) R_{\ell} = 0 \, .
    \label{eq:R_eom}
\end{align}
Note that $R_{\ell} = R_{|\ell|}$, but since our main focus in the main text is on the $\ell = 0$ mode we drop the absolute value for simplicity. Since $\Delta(\rho) = \delta \Theta(1 - \rho)$ is piecewise, to find the solution everywhere we first find solutions for $\rho < 1$ and $\rho > 1$, and then require that $R_{\ell}$ is continuous and differentiable at $\rho = 1$. The solutions to Eq.~\eqref{eq:R_eom} are dramatically different depending on the sign of $\eta$, and therefore we discuss them separately.

\vspace{1em}
\textbf{Case I (Bound Modes): $- \delta < \eta < 0$}. When $- \delta < \eta < 0$ the $R_{\ell}$ solutions in the $\rho < 1$ and $\rho > 1$ regions are,
\begin{align}
    R_{\ell}(\rho; q, \eta) = \begin{cases}
        J_{\ell}( \, \alpha(q, \eta) \, \rho ) & \rho < 1 \\ 
        A_{\ell}(q, \eta) \, K_{\ell}( \, \beta(q, \eta) \, \rho ) & \rho > 1 \, ,
    \end{cases}
    \label{eq:R_bound}
\end{align}
where,
\begin{align}
    \alpha(q, \eta) \equiv \frac{q}{\lambda} \sqrt{2 ( \eta + \delta ) }~~~,~~~\beta(q, \eta) \equiv \frac{q}{\lambda} \sqrt{2 |\eta|}~~~,~~~\chi(q) \equiv \frac{q}{\lambda} \sqrt{2 \delta} \, ,
    \label{eq:alpha_beta_chi_def}
\end{align}
are real constants satisfying $\alpha^2 + \beta^2 = \chi^2$, $J_\ell$ are the Bessel functions of the first kind, and $K_\ell$ are the modified Bessel functions of the second kind. Since $K_\ell(\rho) \rightarrow 0$ as $\rho \rightarrow \infty$, these solutions are ``bound", i.e., they are localized near $\rho < 1$. Continuity and differentiability of $R_{\ell}$ determine the normalization factor, $A_{\ell} = J_\ell(\alpha) / K_\ell(\beta)$, and quantize the $\alpha$, $\beta$ solutions via the following equations,
\begin{align}
    \frac{\alpha \left( J_{\ell + 1}(\alpha) - J_{\ell - 1}(\alpha) \right)}{J_\ell(\alpha)} & = \frac{\beta \left( K_{\ell + 1}(\beta) + K_{\ell - 1}(\beta) \right)}{K_\ell(\beta)}~~~,~~~\alpha^2 + \beta^2 = \chi^2 \, .
    \label{eq:bound_solve}
\end{align}
For a given $q$, the quantization of $\alpha, \beta$ leads to the quantization of $\eta$ using Eq.~\eqref{eq:alpha_beta_chi_def}. There are only a finite number of solutions satisfying Eq.~\eqref{eq:bound_solve}, which we label with $m$. We define the solutions for $\eta$ to be $\eta_{\ell m}(q)$. A visual representation of the $\ell = 0$ solutions is given in Fig.~\ref{fig:app_bound_mode_solutions}. Each intersection of the semi-circle and black lines is a solution, and therefore there are more solutions the larger the radius of the semi-circle, $\chi$, is. 

Given a $q$ we have identified all bound solutions, labeled by $\ell, m$ quantum numbers. The real solutions for the bound mode displacements can be written by combining the $\pm q$, $\pm m$, and $\pm \tau$ solutions discussed previously,
\begin{align}
    u_{\ell m}(\rho, \phi, \zeta, \tau; q) = R_{\ell m}(\rho; q) \begin{Bmatrix} \sin{( \ell \phi )} \\ \cos{( \ell \phi )} \end{Bmatrix} \begin{Bmatrix}
        \sin{(  q (1 - \eta_{\ell m}(q) ) \zeta ) } \\ \cos{( q (1 - \eta_{\ell m}(q) ) \zeta ) } 
    \end{Bmatrix} \begin{Bmatrix}
        \sin{( q \tau ) } \\ \cos{( q \tau ) } 
    \end{Bmatrix} \, ,
    \label{eq:all_bound_u}
\end{align}
where the $\ell m$ subscripts on $u$ and $R$ indicate that they are evaluated at $\eta = \eta_{\ell m}(q)$, e.g. $R_{\ell m}(\rho; q) \equiv R_{\ell}(\rho; q, \eta_{\ell m}(q))$. 

The last undetermined quantity is $q$, which is set by satisfying Eq.~\eqref{eq:z_boundary_cartesian}. This picks out the $\cos{(q ( 1 - \eta) \zeta )}$ solution of Eq.~\eqref{eq:all_bound_u}, and must satisfy the quantization condition, $q (1 - \eta_{\ell m}(q)) = n \pi$. Since $\eta \ll 1$ the leading order solutions are simply $q_{n \ell m} = n \pi$. The next order solutions are found by solving,
\begin{align}
    q_{n \ell m} (1 - \eta_{n \ell m}) \approx \pi n~~~\implies~~~q_{n \ell m}  \approx \pi n (1 + \eta_{n \ell m}) \, ,
\end{align}
where $\eta_{n \ell m} \equiv \eta_{\ell m}( n \pi )$. Returning to dimensionful coordinates the bound mode solutions are,
\begin{align}
    u_{n\ell m}(r, \phi, z, t) = R_{n \ell m}(r / R) \cos{\left( \frac{n \pi z}{L + h} \right)} \begin{Bmatrix} \sin{( \ell \phi )} \\ \cos{( \ell \phi )} \end{Bmatrix} \begin{Bmatrix}
        \sin{( \omega_{n\ell m} t ) } \\ \cos{( \omega_{n \ell m} t ) }
    \end{Bmatrix}~~~,~~~\omega_{n \ell m} \approx \frac{ n \pi c_\text{l}}{L} \left( 1 + \eta_{n \ell m} \right) \, ,
    \label{eq:bound_u_final}
\end{align}
where the $n, \ell, m$ subscripts on $u$ and $R$ indicate that they are evaluated at $q = q_{n \ell m}$ and $\eta = \eta_{n \ell m}$. The $\ell = 0$ mode functions from Eq.~\eqref{eq:bound_u_final} are also given in Eq.~\eqref{eq:displacement} in the main text.

\vspace{1em}
\textbf{Case II (``Free" Modes): $\eta > 0$}. When $\eta > 0$ the solutions for $\rho > 1$ no longer decay; instead they are a linear combination of $J_\ell$ and $Y_\ell$, where the latter are the Bessel functions of the second kind,
\begin{align}
    R_{\ell}(\rho; q, \eta) = \begin{cases}
        J_{\ell}( \alpha(q, \eta) \, \rho ) & \rho < 1 \\ 
        A_{\ell}(q, \eta) \, J_{\ell}( \beta(q, \eta) \, \rho ) + B_{\ell}(q, \eta) \, Y_{\ell}( \beta(q, \eta) \, \rho ) & \rho > 1 \, .
    \end{cases}
    \label{eq:R_free}
\end{align}
We refer to these solutions as ``free" since as $\rho\rightarrow \infty$ both $J_\ell$ and $Y_\ell$ become a linear combination of sin and cos functions. For these modes, continuity and differentiability of $R_{\ell}$ at $\rho = 1$ is enough to determine $A_{\ell}, B_{\ell}$,
\begin{align}
     A_{\ell} = \frac{1}{2} \pi  (\beta  J_\ell(\alpha ) Y_{\ell-1}(\beta )-\alpha  J_{\ell-1}(\alpha ) Y_\ell(\beta ))~~~,~~~B_{\ell} = -\frac{1}{2} \pi  (\beta  J_\ell(\alpha ) J_{\ell-1}(\beta )-\alpha  J_{\ell-1}(\alpha ) J_\ell(\beta ))\, ,
\end{align}
but, opposite of the $\eta < 0$ solutions, $\alpha, \beta$ are \textit{not} quantized. The quantization of these modes will depend on the boundary conditions of the target chip in the directions perpendicular to $\hat{\vec{z}}$, but since we do not include these conditions the radial modes form a continuum. However the boundary condition in Eq.~\eqref{eq:z_boundary_cartesian} must still be satisfied. This is done in the same way as the $\eta < 0$ case and results in,
\begin{align}
    q_{n}(\eta) (1 - \eta) = \pi n~~~\implies~~~q_{n}(\eta)  \approx \pi n (1 + \eta) \, .
\end{align}

To summarize, without enforcing boundary conditions on $R_{\ell}$, any $\eta > 0$ (as long as $\eta \ll 1$) is a solution. For any choice of $\eta$ there are modes labeled by $n, \ell$ whose real, dimensionful, phonon mode functions are,
\begin{align}
    u_{n \ell}(r, \phi, z, t; \eta) = R_{n \ell}(r / R; \eta) \cos{\left( \frac{n \pi z}{L + h} \right)} \begin{Bmatrix} \sin{( \ell \phi )} \\ \cos{( \ell \phi )} \end{Bmatrix} \begin{Bmatrix}
        \sin{( \omega_{n}(\eta) \, t ) } \\ \cos{( \omega_{n}(\eta) \, t ) }
    \end{Bmatrix}~~~,~~~\omega_{n}(\eta) \approx \frac{ n \pi c_\text{l}}{L} \left( 1 + \eta \right) \, ,
    \label{eq:bound_u_final_2}
\end{align}
where the $n, \ell$ subscripts on $u$ and $R$ indicate that they are evaluated at $q = q_n(\eta)$, and, as in the $\eta < 0$ case, the real solutions have been constructed out of a linear combination of the exponential solutions.

\section{Backgrounds}
\label{app:backgrounds}

Here we further discuss the dark count backgrounds, focusing on three topics: in Sec.~\ref{subsubsec:DiscussionOfGeneralBackgroundModel} we discuss our model of the background, in Sec.~\ref{subsec:CommentsOnDarkCountComponents} we discuss the assumptions underlying the estimate for the spontaneous excitation rates in Eqs.~\eqref{eq:p_exc_th} and~\eqref{eq:p_exc_qp}, and lastly in Sec.~\ref{subsec:ModelForDeltaPdc} we discuss estimates of the magnitude of the systematic uncertainty, $\delta p_\text{dc}$ in Eq.~\eqref{eq:psmin}.

\subsection{The Background Model}
\label{subsubsec:DiscussionOfGeneralBackgroundModel}

\begin{figure}
    \centering
    \includegraphics[width=0.9\linewidth]{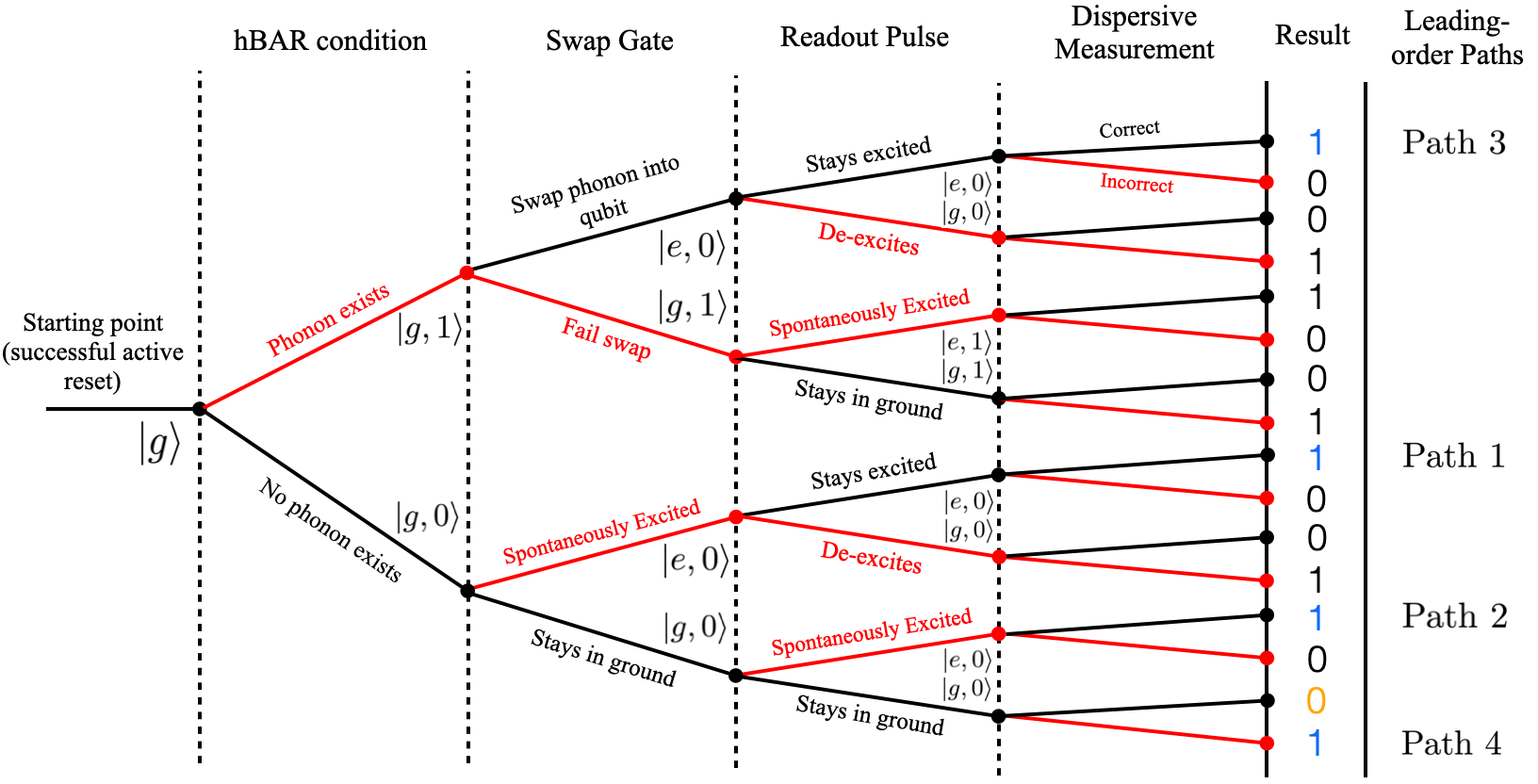}
    \caption{A simplified diagram of the potential outcomes of an individual measurement, with their respective measurement results. Paths whose result is a 1 constitute a background, whereas those with a 0 do not. Each node/fork represents two outcomes for a particular step in the measurement process. Red segments indicate steps with ``small'' probabilities: a failed swap has a probability $1-\varepsilon_{\text{s}}$, (with $\varepsilon_{\text{s}}\sim\mathcal{O}(1)$), de/excitation probabilities are small due to small window lengths, and good readout fidelities make a readout failure unlikely. Black segments indicate steps with $\mathcal{O}(1)$ probabilities. The total probability of each path from the starting point to the result is given by the product of the probabilities of all steps along that path. Paths 1-4, for which the result is colored blue, only have a single red path segment, and are therefore the leading-order components of this background model. The result colored orange is the only path with $\mathcal{O}(1)$ probability, occurs when there are no spurious excitations or readout failures, and does not contribute background counts.}   \label{fig:app_branches_of_a_measurement}
\end{figure}

Successfully measuring the state of the $h$BAR requires a number of steps, each of which can fail and lead to dark counts. These steps are illustrated in Fig.~\ref{fig:app_branches_of_a_measurement}, where each of the possible outcomes are represented in a branching diagram. The total probability that any path is followed is given by the product of the probabilities at each step, and the total probability of a dark count is the sum of all paths which lead to a measured ``1". Since the probabilities of each step are asymmetric, i.e., one path is much more likely than the other, the set of paths with only one unlikely step dominate the backgrounds. This is illustrated in Fig.~\ref{fig:app_branches_of_a_measurement} where the more likely path at each node is black, and the less likely path is red. The set of paths with only one red segment are the leading order background contributions. These paths are labelled ``Paths 1-4" and have probabilities $p_{1-4}$, respectively. Note that while the probability of each path will not correspond to each term in Eq.~\eqref{eq:p_dc}, the sum of all $p_{1-4}$ will match the sum in Eq.~\eqref{eq:p_dc}, i.e., $p_1 + p_2 + p_3 + p_4 = p_{\text{th},h} + p_{\text{th,q}} + p_{\text{qp}} + p_{\text{ro}}$. We now discuss what goes wrong in each path.

\begin{figure}
    \centering
    \includegraphics[width=0.95\linewidth]{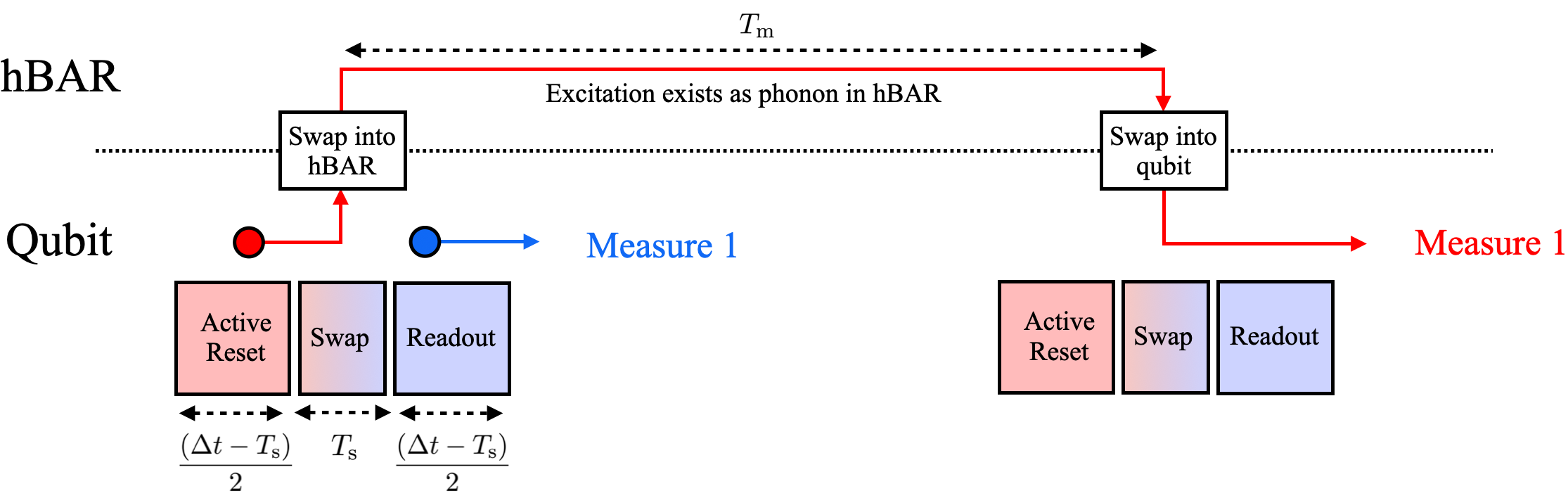}
    \caption{A diagram of two measurement sequences, showing how excitations (circles) at different points of the reset-swap-readout scheme will either yield a dark count in the current measurement (blue), or a potential dark count in a future measurement of the same mode due to injected phonons (red).}
    \label{fig:app_injected_phonons}
\end{figure}

Paths 1 and 2 are from spontaneous qubit excitations after the active reset. These can be due to both thermal qubit excitations and quasiparticle (QP)-induced qubit excitations, as discussed in Sec.~\ref{subsec:Backgrounds}, and the sum of the probabilities is,
\begin{equation}
    \label{eq:p_mid}
    p_{1} + p_{2} \approx \frac{\Delta t}{2}\Bigg( \Bigg[\frac{\, \omega_\text{q}}{Q_\text{q}} \, \frac{1}{e^{\omega_\text{q}/T_{\text{q}}^{\text{eff}}} - 1}\Bigg] + \Bigg[2.17 \Bigg(\frac{\Delta}{\omega_\text{q}}\Bigg)^{3.65} \sqrt{\frac{2\omega_\text{q}\Delta}{\pi^{2}}} x_\text{qp}^{2}\Bigg]\Bigg) \, .
\end{equation}
The terms inside the parentheses are the excitation rates from thermal and QP-induced qubit excitations, respectively. These are multiplied by $\Delta t / 2$ since only excitations in the second half of a measurement can affect this measurement cycle (this is illustrated by the blue dot and arrow in Fig.~\ref{fig:app_injected_phonons} and will be discussed shortly).

Path 3 is from successfully detecting a real background phonon in the $h$BAR, either from thermal production in the $h$BAR, or from errors in the first half of the \textit{previous} measurement cycle, (e.g., from active reset) which were swapped into the $h$BAR and subsequently swapped back into the qubit during the current measurement cycle. The thermal contribution from the $h$BAR is simply $p_{\text{th},h}$ in Eq.~\eqref{eq:p_dc}. The probability of an error in the first half of the previous measurement is the same as Eq.~\eqref{eq:p_mid}. However, these phonons must exist until the current measurement, a time $T_\text{m}$ later, and there is some probability they decay. Therefore,
\begin{align}
    p_3 = p_{\text{th,h}} + (p_1 + p_2) \, e^{- T_\text{m} \omega_\text{p} / Q_\text{p}} \, .
\end{align}
Fig.~\ref{fig:app_injected_phonons} further conceptually illustrates the distinction between the spurious excitation backgrounds in the two halves of a measurement cycle, and clarifies how excitations occurring in the first half of a cycle end up affecting the following cycle via a phonon inadvertently ``injected'' into the $h$BAR.

Lastly, Path 4 is a readout failure when there are no true excitations, and therefore approximately \(p_{4} \approx p_{\text{ro}} \), where $p_\text{ro} = (1 - \mathcal{F})/2$ and $\mathcal{F}$ is the ideal single shot fidelity, as discussed in Sec.~\ref{subsec:Backgrounds}.

\subsection{Spontaneous Qubit Excitation Rate}
\label{subsec:CommentsOnDarkCountComponents}

We now discuss the spontaneous qubit excitation rate, starting with a discussion of the QP-induced excitations, and then moving to the thermal excitations. Our models for QP-induced qubit excitation are based on Ref.~\cite{WennerHotQuasiparticles}, which presents an estimate for the steady-state excited-state probability \(p_{e}\) in terms of the reduced QP density \(x_\text{qp} \equiv n_\text{qp}/n_\text{CP}\), where \(n_\text{qp}\) is the volume density of QPs and \(n_\text{CP}\) is the volume density of Cooper pairs. Assuming \(p_{e} \ll 1\), this  work estimates the excited-state probability \(p_{e}\) as:
\begin{equation}
\label{eq:WennerEq}
p_{e} = \frac{\Gamma_{\uparrow,\text{qp}}}{\Gamma_{\downarrow,\text{qp}}+\Gamma_{\uparrow,\text{qp}}} = 2.17 \Bigg(\frac{\Delta}{\hbar\omega_\text{q}}\Bigg)^{3.65} x_\text{qp}.
\end{equation}
This functional form is a fit to an \(x_\text{qp}\)-dependent calculation of \(p_{e}\) that critically depends on the energy distribution of the QPs in the qubit junction leads. Using this, we may find that for a transmon-style qubit,
\begin{align}
\label{eq:GammaUp2}
\Gamma_{\uparrow,qp} \simeq p_{e}\Gamma_{\downarrow,qp} =&\ 2.17 \Bigg(\frac{\Delta}{\omega_\text{q}}\Bigg)^{3.65} \sqrt{\frac{2\omega_\text{q}\Delta}{\pi^{2}\hbar}} x_\text{qp}^{2}
\end{align}
where a standard ``cold-QP'' downward transition rate is used for \(\Gamma_{\downarrow,\text{qp}}\), as suggested in Ref.~\cite{CatelaniRelaxation,McEwen}. This rate is used as input for the excitation probability in the main text: for a window \(\Delta t\) during which excitations can occur, the excitation probability is modeled as  \(p_{\text{qp}}=\Gamma_{\uparrow,\text{qp}}\Delta t\).

Ref.~\cite{WennerHotQuasiparticles} estimates Eq.~\eqref{eq:WennerEq} by modeling the QP energy distribution under the assumption that QPs are injected directly into the superconducting layer. While this may be somewhat representative of transient processes such as in-substrate energy depositions from cosmic rays that result in phonon-mediated QP production, it does not capture all processes that may cause QP-associated qubit excitation. For example other experiments, e.g., Ref.~\cite{Serniak2018}, demonstrated significantly higher ratios of QP-induced upward transition rates to QP-induced downward transition rates than those predicted by this functional form. While such observations could be attributed to a ``hot'' QP population with characteristic energy not localized to near the gap \(\Delta\), photon-assisted qubit excitation processes may also explain these excitation rates~\cite{HouzetPAPS}. In principle, this source of excitations should be modeled and may have a different functional form than that in Eq~\eqref{eq:WennerEq}. However, as recent experiments have claimed that photon-assisted processes do not limit their excited state population~\cite{ConnollyYale2024}, we make the simplifying assumption that engineering strategies like radiation shielding of environment and input/output lines may make this source of backgrounds subdominant. We therefore use the functional form in Eq.~\eqref{eq:GammaUp2} to estimate our dark count contributions from QP-induced excitations, recognizing that there are uncertainties in modeling the QP energy distribution in the junction leads for arbitrary QP injection sources.

Spontaneous qubit excitations may also occur from qubit interaction with a non-QP thermal bath, similar to that formed from a set of nearby two-level systems that are able to accept and donate energy to the qubit~\cite{LuHeatTransport}. While the qubit and \hbarName{} are thermalized to the mixing chamber plate of a fridge at temperatures typically around 10-20~mK, the qubit itself is often locally heated well above this temperature, by, for example, hot quasiparticles and readout (RF) noise~\cite{LuHeatTransport}. Moreover, the effective temperatures of the QP population and non-QP dissipative baths can often be different~\cite{Serniak2018}, illustrating a degree of decoupling between these sources of excitations.  In our model of non-QP thermal bath interactions, we parameterize this bath as having an effective temperature \(T_{\text{q}}^{\text{eff}}\). We also assume the thermal bath follows the Markov approximation, i.e. that correlations within the bath dissipate over timescales much smaller than experimentally relevant timescales (here, the inter-measurement time). This approximation is needed in order to assume independence between results of the target and sideband measurements in a given measurement set. While experimental probes have demonstrated both Markovian and non-Markovian dynamics~\cite{NonMarkovianPhononicBandgap,ShaliboTLS,BurnettDecoherenceBenchmarking,MullerTLSReview} in superconducting qubits coupled to a TLS bath, we assume that non-Markovian dynamics from nearby, long-lived TLSs may be ``calibrated out'' by exploiting common TLS properties such as dependence on applied strain~\cite{Lisenfeld_2019,Lisenfeld_2016}. Further experimental probes are needed to confirm this capability, and are planned for pathfinder/prototype devices.

\subsection{Systematic Dark Count Rate Fluctuations}
\label{subsec:ModelForDeltaPdc}

Systematic dark counts can be generated by short timescale, $\mathcal{O}(10 \, \mu\text{s})$, fluctuations in the system parameters. While estimates exist for longer timescales, short timescale fluctuations have not been studied. However as a rough estimate it is useful to propagate measured fluctuations to $\delta p_\text{dc}$, given the background model discussed in Sec.~\ref{subsec:Backgrounds}. We limit ourselves to considering fluctuations of the qubit lifetime~\cite{KlimovFluctuations,Muller2015Fluctuations,BurnettDecoherenceBenchmarking} and qubit frequency~\cite{KlimovFluctuations,SchlorFluctuations}, though additional parameters such as the QP density \(x_\text{qp}\), effective qubit and \hbarName{} temperatures, and amplifier gain (which affects ideal single shot fidelity \(\mathcal{F}\)) will also fluctuate. Following the prescription in Sec.~\ref{subsec:Backgrounds} and focusing on variations in the qubit frequency (with \(\delta \omega_\text{q}/\omega_\text{q}=500\, \text{kHz}/4.75 \,\text{GHz}\)~\cite{KlimovFluctuations}) and qubit lifetime (with \(\delta\tau_\text{q}/\tau_\text{q}\simeq0.1\)~\cite{BurnettDecoherenceBenchmarking}), we estimate \(\delta p_{\text{dc}}/p_{\text{dc}}\simeq 0.03\), motivating a conservative choice of \(\delta p_{\text{dc}}/p_{\text{dc}}\lesssim0.1\) for the prototype development stage. 

\section{Proposed Experimental Setup}
\label{app:ExperimentalSetup}

\begin{figure}
    \centering
    \includegraphics[width=0.6\linewidth]{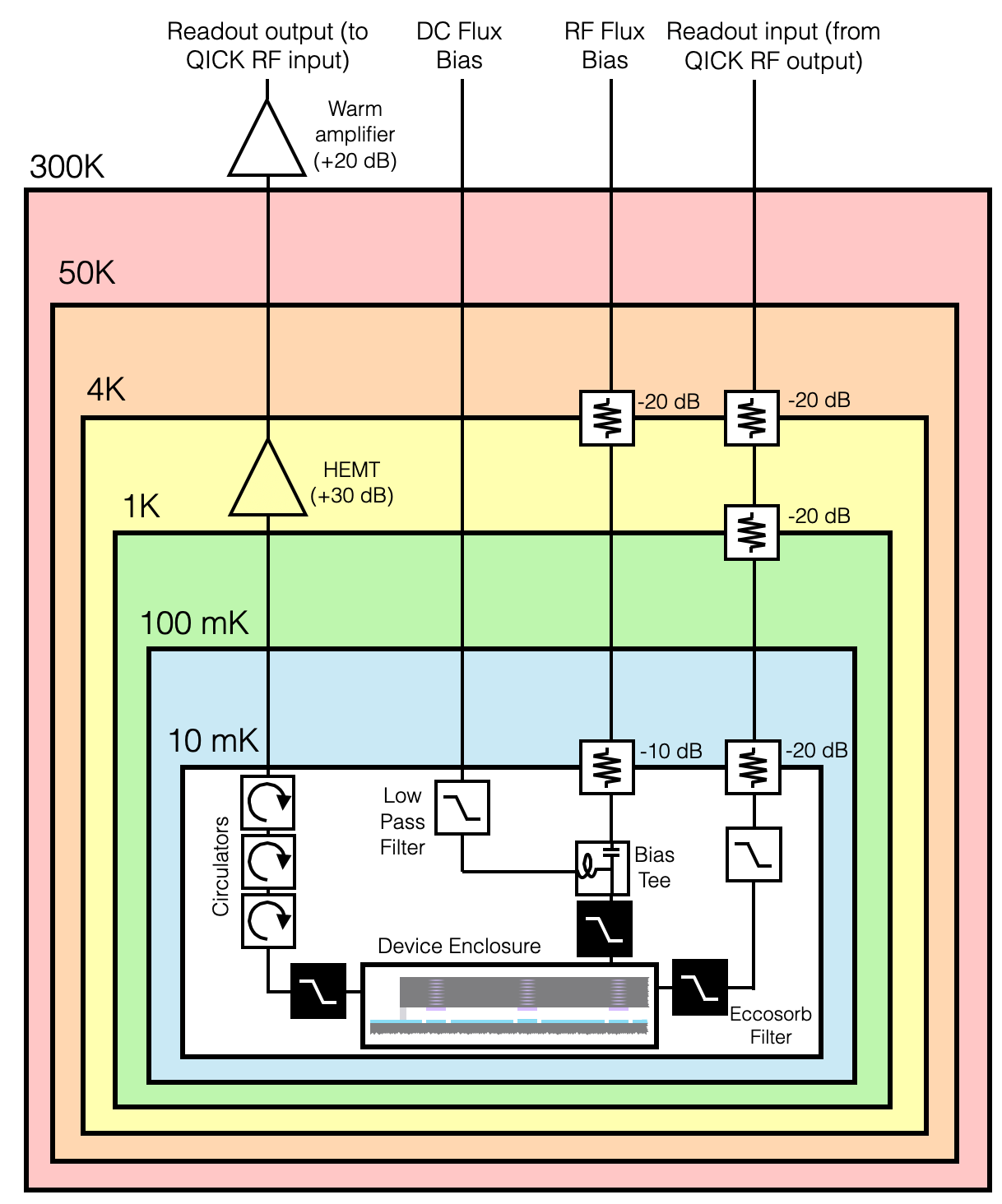}
    \caption{Proposed experimental configuration for the qc-$h$BAR operation within a standard dilution refrigerator environment.}
    \label{fig:app_proposed_experimental_setup}
\end{figure}

In Fig.~\ref{fig:app_proposed_experimental_setup} we present a possible experimental configuration for the qc-$h$BAR device. The device is thermally anchored to the base plate of a dilution refrigerator, which typically reaches 10 mK. In our simplified diagram, the RF signal for probing the state of the qubit originates from a QICK board's RF output~\cite{QICK_2022} and passes onto the readout input line shown. The signal passes through a set of attenuators which balance the thermal noise at the baseplate with the power dissipated in each plate, and then passes into the device. The output of the dispersive measurement passes through a set of circulators to limit reflections, and up through cold (HEMT) and warm amplification before being sent into the QICK board’s RF input chain for digitization. We also show two magnetic flux bias lines here: one DC flux bias for slow control of the qubit frequency \(\omega_{\text{q}}\) with applied magnetic flux, and one RF flux bias for the fast operations involved in moving the qubit frequency during the swap gate (though this may also be achieved through a Stark shift). Other critical elements are the eccosorb filters, which limit IR radiation that may break Cooper pairs or otherwise thermally heat the qubit to increase the background probabilities $p_{\text{qp}}$ or $p_{\text{th,q}}$. While we do not include parametric amplification in this diagram, it may also help to improve single-shot fidelity, a dominant source of backgrounds in much of the $1<\omega_{\text{q}}/2\pi<10$~GHz frequency range. 

\bibliographystyle{utphys3}
\bibliography{bibliography}

\end{document}